\documentclass[]{spie}  
\usepackage[utf8x]{inputenc}

\usepackage{amsmath,amsfonts,amssymb}
\usepackage{graphicx}
\usepackage[colorlinks=true, allcolors=blue]{hyperref}
\usepackage[table]{xcolor}
\usepackage{graphicx}
\usepackage{txfonts}
\usepackage{lscape}
\usepackage{hyperref}
\usepackage{comment}
\usepackage{xspace}

\usepackage{bm}
\usepackage{aas_macros}

\renewcommand\vec{\bm}

\def \us{$\mu s$\xspace}

\title{TIMING TECHNIQUES APPLIED TO DISTRIBUTED MODULAR HIGH-ENERGY ASTRONOMY: THE HERMES PROJECT}

\author[a]{A.~Sanna}
\author[b]{A.~F.~Gambino}
\author[a]{L.~Burderi}
\author[a]{A.~Riggio}
\author[b]{T.~Di Salvo}
\author[c]{F.~Fiore}
\author[d]{M.~Lavagna}
\author[e]{R.~Bertacin}
\author[f,$\gamma$]{Y.~Evangelista}
\author[g]{R.~Campana}
\author[g]{F.~Fuschino}
\author[d]{P.~Lunghi}
\author[h]{A.~Monge}
\author[e]{B.~Negri}
\author[e]{S.~Pirrotta}
\author[e]{S.~Puccetti}
\author[i]{F.~Amarilli}
\author[f]{F.~Ambrosino}
\author[j,k]{G.~Amelino-Camelia}
\author[a]{A.~Anitra}
\author[a]{M.~Barbera}
\author[d]{M.~Bechini}
\author[l]{P.~Bellutti}
\author[m]{G.~Bertuccio}
\author[n]{J.~Cao}
\author[f]{F.~Ceraudo}
\author[n]{T.~Chen}
\author[o]{M.~Cinelli}
\author[p]{M.~Citossi}
\author[q]{A.~Clerici}
\author[d]{A.~Colagrossi}
\author[d]{S.~Curzel}
\author[p]{G.~Della Casa}
\author[l]{E.~Demenev}
\author[r]{M.~Del Santo}
\author[p]{G.~Dilillo}
\author[s]{P.~Efremov}
\author[f,$\gamma$]{M.~Feroci}
\author[c]{C.~Feruglio}
\author[m]{F.~Ferrandi}
\author[t]{M.~Fiorini}
\author[d]{M.~Fiorito}
\author[s]{D.~Gacnik}
\author[u]{G.~Galg\'oczi}
\author[n]{N.~Gao}
\author[m]{M.~Gandola}
\author[v]{G.~Ghirlanda}
\author[s]{A.~Gomboc}
\author[w]{M.~Grassi}
\author[$\psi$]{C.~Guidorzi}
\author[x]{A.~Guzman}
\author[b]{R.~Iaria}
\author[s]{M.~Karlica}
\author[q]{U.~Kostic}
\author[g]{C.~Labanti}
\author[r]{G.~La Rosa}
\author[r]{U.~Lo Cicero}
\author[h]{B.~Lopez Fernandez}
\author[w]{P.~Malcovati}
\author[e,$\beta$]{A.~Maselli}
\author[a]{A.~Manca}
\author[m]{F.~Mele}
\author[y]{D.~Mil\'ankovich}
\author[g]{G.~Morgante}
\author[v]{L.~Nava}
\author[r]{P.~Nogara}
\author[z]{M.~Ohno}
\author[d]{D.~Ottolina}
\author[d]{A.~Pasquale}
\author[$\alpha$]{A.~Pal}
\author[e,$\beta$]{M.~Perri}
\author[d]{M.~Piccinin}
\author[f]{R.~Piazzolla}
\author[u]{S.~Pliego-Caballero}
\author[d]{J.~Prinetto}
\author[o]{G.~Pucacco}
\author[$\delta$]{I.~Rashevskaya}
\author[$\delta$]{A.~Rashevski}
\author[z,$\epsilon$]{J.~Ripa}
\author[r]{F.~Russo}
\author[$\beta$]{A. Papitto}
\author[$\beta$]{S.~Piranomonte}
\author[x]{A.~Santangelo}
\author[d]{F.~Scala}
\author[f]{G.~Sciarrone}
\author[u]{D.~Selcan}
\author[d]{S.~Silvestrini}
\author[r]{G.~Sottile}
\author[u]{T.~Rotovnik}
\author[x]{C.~Tenzer}
\author[d]{I.~Troisi}
\author[p,$\zeta$]{A.~Vacchi}
\author[g]{E.~Virgilli}
\author[z,$\epsilon$]{N.~Werner}
\author[n]{L.~Wang}
\author[n]{Y.~Xu}
\author[$\eta$]{G.~Zampa}
\author[$\eta$, $\zeta$]{N.~Zampa}
\author[d]{G.~Zanotti}

\affil[a]{Dipartimento di Fisica, Universit\`a degli Studi di Cagliari, SP Monserrato-Sestu km 0.7, I-09042 Monserrato, Italy}
\affil[b]{Dipartimento di Fisica e Chimica, Universit\`a degli Studi di Palermo, via Archirafi 36, I-90123 Palermo, Italy}
\affil[c]{INAF-OATS, Via G.B. Tiepolo 11, I-34143, Trieste, Italy}
\affil[d]{Politecnico di Milano, Via La Masa 34, 20156, Milano, Italy}
\affil[e]{Agenzia Spaziale Italiana, via del Politecnico snc, 00133 Roma, Italy}
\affil[f]{INAF-IAPS Rome, Via del Fosso del Cavaliere 100, I-00133, Italy}
\affil[g]{INAF-OAS Bologna, Via Gobetti 101, I-40129, Bologna, Italy}
\affil[h]{DEIMOS, Spain}
\affil[i]{Fondazione Politecnico di Milano, Piazza Leonardo da Vinci, 32 20133 Milano, Italy}
\affil[j]{Dipartimento di Fisica Ettore Pancini, Universit\`a di Napoli Federico II}
\affil[k]{INFN, Sezione di Napoli, Complesso Univ. Monte S. Angelo, I-80126 Napoli, Italy}
\affil[l]{Fondazione Bruno Kessler - FBK, Via Sommarive 18, I-38123 Trento, Italy}
\affil[m]{Department of Electronics, Information and Bioengineering (DEIB) of Politecnico di Milano, Como Campus, Via Anzani 42, 22100 Como, Italy}
\affil[n]{Institute of High Energy Physics, Chinese Academy of Sciences, China}
\affil[o]{Dipartimento di Matematica, Universit\`a di Roma Tor Vergata}
\affil[p]{Universit\`a degli Studi di Udine, Via delle Scienze, 206, 33100 Udine, Italy}
\affil[q]{Aalta Lab, Slovenia}
\affil[r]{INAF-IASF Palermo, Via U. La Malfa 153, I-90146 Palermo, Italy}
\affil[s]{University of Nova Gorica, Slovenia}
\affil[t]{INAF-IASF Milano, Via Bassini 15, I-20100 Milano, Italy}
\affil[u]{Skylabs, Slovenia}
\affil[v]{INAF-OAB,Via E. Bianchi 46, I-23807 Merate, Italy}
\affil[w]{University of Pavia, Department of Electrical, Computer, and Biomedical Engineering, Via Ferrata 5, I-27100, Pavia, Italy}
\affil[x]{IAAT University of Tuebingen, Sand 1 - 72076 Tuebingen, Germany}
\affil[y]{C3S, Hungary}
\affil[z]{ELTE - E\"ot\"ovs Lor\'and University, Hungary}
\affil[$\alpha$]{Konkoly Observatory, Hungary}
\affil[$\beta$]{INAF - Osservatorio Astronomico di Roma, via Frascati 33, I-00040 Monteporzio Catone, Italy}
\affil[$\gamma$]{INFN}
\affil[$\delta$]{TIFPA-INFN}
\affil[$\epsilon$]{Department of Theoretical Physics and Astrophysics, Masaryk University, Brno, Czech Republic}
\affil[$\zeta$]{INFN Udine, Via delle Scienze 206, I-33100 Udine, Italy}
\affil[$\eta$]{INFN sez. Trieste, Padriciano 99, I-34127 Trieste, Italy}
\affil[$\psi$]{Dipartimento di Fisica e scienze della Terra, Universit\`a di Ferrara, Italy}
\authorinfo{Further author information: \\Andrea Sanna: E-mail: andrea.sanna@dsf.unica.it, hermes-sp.eu}

\begin{document} 
\maketitle

\begin{abstract}
The HERMES-TP/SP (High Energy Rapid Modular Ensemble of Satellites - Technologic and Scientific Pathfinder) is an in-orbit demonstration of the so-called distributed astronomy concept. Conceived as a mini-constellation of six 3U nano-satellites hosting a new miniaturized detector, HERMES-TP/SP aims at the detection and accurate localisation of bright high-energy transients such as Gamma-Ray Bursts. The large energy band, the excellent temporal resolution and the wide field of view that characterize the detectors of the constellation represent the key features for the next generation high-energy all-sky monitor with good localisation capabilities that will play a pivotal role in the future of \emph{Multi-messenger Astronomy}. 
In this work, we will describe in detail the temporal techniques that allow the localisation of bright transient events taking advantage of their almost simultaneous observation by spatially spaced detectors. Moreover, we will quantitatively discuss the all-sky monitor capabilities of the HERMES Pathfinder as well as its achievable accuracies on the localisation of the detected Gamma-Ray Bursts. 
\end{abstract}


\section{INTRODUCTION}

The HERMES (High Energy Rapid Modular Ensemble of Satellites) project is based on a revolutionary mission concept also known as distributed astronomy. More specifically, HERMES is conceived as a constellation of CubeSats distributed in low Earth orbits and hosting technologically advanced X/gamma-ray detectors, with relatively small effective area, that aim at localise and investigate the spectral and temporal properties of bright highly-energetic transient events.

Interestingly, we are now in the middle of a transitional phase both scientifically and technologically speaking. On the one hand, we recently witnessed the beginning of the so-called \emph{Multi-messenger Astronomy}, that coincided with the observation of the Gravitational Wave Event GW170817 by the Advanced LIGO/Virgo [\citenum{GW170817_GW}], followed by the detection of the associated short Gamma-Ray Burst (GRB 170817A) detected by the Fermi and INTEGRAL satellites [\citenum{Troja17}], as well as the countless follow-up multi-wavelength observations [see e.g. \citenum{Cowperthwaite17,Siebert17}]. Immediately after this discovery, it became even clearer for the community the urgency for an high-energy all-sky monitor with good localisation capabilities to work in parallel with gravitational wave observatories such as Ligo/Virgo/Kagra that will reach their final sensitivity within a few years. At the moment, all the available satellites providing the detection and localisation of GRBs have been operational for more than a decade with large chances of decommissioning in the next years, while next-generation large-area detectors will not be available for at least 10-15 years from now. 

On the other hand, we are living on the verge of a major revolution where new technologies can finally challenge the current assumptions that high-energy astronomical observations from space can only be achieved by big satellites, usually designed, built, launched and managed by government space agencies. CubeSats, light spacecrafts with small sizes and reduced costs, are now considered a competitive solution for space applications as they allow equilibrium among crucial variables of a space project, such as development time, cost, reliability, mission lifetime, and replacement [see e.g. \citenum{douglas19, Shkolnik18}]. Interestingly, they represent the key to realize the revolutionary concept of distributed (modular) space astronomy crucial to investigate the unknown of the Universe. Indeed, the large number of photons required to perform cutting edge (astro)physical space science can be collected adding up the contribution of a large number of detectors distributed over a fleet of nano/micro/small-satellites, allowing to achieve huge overall collecting areas which otherwise would be unreachable with a single instrument and impossible to carry with the current rocket load capabilities. The decreasing trend of producing costs as well as the increase of launching opportunities witnessed in recent years, allows us to concretely conceive fleets of hundreds/thousands of coordinated satellites acting as a single detector of unprecedented collecting power.

The full concept of HERMES, also known as the HERMES Full Constellation (HFC) has been conceived to tackle three main scientific objectives: 
\begin{itemize}
\item the accurate and prompt localization of bright X/gamma-ray transients such as GRBs. Fast high-energy transients are among the likely electromagnetic counterparts of the gravitational wave events (GWE) recently discovered by Advanced LIGO/Virgo, and of the Fast Radio Burst; 
\item Open a new window on timing at X/gamma-ray energies, and thus investigate for the first time the intrinsic temporal variability of GRBs down to fractions of micro-seconds, to constrain models for the GRB engine;
\item Test quantum space-time scenarios by measuring the delay time between GRB photons at different energies.
\end{itemize}

The determination of the position in the sky of an astrophysical transient source is crucial to investigate its origin. A clear example of that is the case of GRBs. Indeed, for almost thirty years after their discovery, GRBs remained poorly understood due to the lack of a precise localization. Only in the late 90s, with the launch of the Italian-Dutch X-ray satellite BeppoSAX [\citenum{Boella97}], the detection of the GRB X-ray afterglow [see e.g. \citenum{vanParadijs97}] and the subsequent identification of the optical and radio transient by ground-based telescopes, allowed the identification of the host galaxy and hence its extragalactic origin (see e.g. [\citenum{Schilling02}] for a review). This paramount discovery allowed to gauge the energy output of GRBs, establishing their cosmological nature. GRBs proved to be, eventually, the most powerful electromagnetic explosions in the Universe, one of the best tools to investigate the Universe during its infancy. Similarly, the recent confirmation that short GRBs are the electromagnetic counterparts of Gravitational Wave Events (see e.g. [\citenum{GW170817_GW,Troja17}]) made the improvement of localisation capabilities of the astronomical observatories a short term technological priority. Indeed, the key to exploiting the \emph{Multi-messenger Astronomy} is the fast and accurate identification of the electromagnetic counterparts of GWEs, crucial to promptly characterize the properties of these events. 

The improved sensibilities of the next generation interferometers will allow the detection of fainter and further away events, making more challenging the identification of their electromagnetic counterparts given the larger portions of space to be searched. Taking as a reference the famous GW170817 event, the horizon for neutron stars merging events detected with a similar signal-to-noise ratio will reach up to 200 Mpc for LIGO and 100-130 Mpc for Virgo in a few years from now. This implies an increase on the discovery volume by a factor $\sim$100 with respect to the GW170817 case. In this scenario, the operation of an efficient X/gamma-ray all-sky-monitor with good localization capability will have a pivotal role in quickly discovering the high-energy counterparts of GWEs and triggering coordinate multi-wavelength observational campaigns.

Another key aspect of modular astronomy is the possibility to combine together the information collected by single elements of the constellation to increase the signal-to-noise ratio, recreating the capabilities of a single observatory with large collecting area. Indeed, once the transients are detected and localized, the signals received by the different detectors can be combined together, after correcting for the delay time of arrivals, significantly increasing their statistics and thus the sensitivity to finer temporal structures. This aspect is particularly important for events such as GRBs,  characterized by huge luminosities and fast variability, often as short as one millisecond. Best most accurate available description of these events is included in the so-called \emph{fireball model}, i.e. a relativistic bulk flow where shocks efficiently accelerate particles (see e.g. [\citenum{Piran99}]). Interestingly, while successful in explaining GRB observations, this model implies a thick photosphere, hampering direct observations of the hidden inner engine that accelerate the bulk flow. One possibility to shed light on their inner engines is through GRB fast variability (see e.g. [\citenum{Kobayashi97, Nakar02, Morsony2010}]). GRB light-curves have been investigated in detail down to $\sim$1 ms timescales or slightly lower (see e.g. [\citenum{Walker00,MacLachlan13}]), while the $\mu$s-ms window is basically unexplored, as poorly little is known also regarding the real duration of the prompt event. It is still unclear how many shells are ejected from the central engine, which is the frequency of ejection and what is its length. With HFC it will be possible to access the $\mu$s-ms timing window for GRBs, allowing for the first time to further investigate their central engines.

The extraordinary capabilities of the HFC will also allow the first dedicated experiment for testing quantum gravity theories. The experiment is founded on the predictions of a discrete structure for space on small scales (of the order of the Planck length) proposed by several theories. This space discretization implies the onset of a dispersion relation for photons, as well as an energy dependence of their propagation speed. A promising method for constraining a first order dispersion relation for photons in vacuo is the study of discrepancies in the arrival times of high-energy photons of GRBs emitted at cosmological distances in different energy bands (see e.g. [\citenum{Burderi20,Burderi20esa}] and references therein for more details on the topic).

Taking advantage of the modularity of the project, the HERMES concept will be tested following a step-by-step strategy. More specifically, at first it will be realized the HERMES Pathfinder with the aim at proving in space the HERMES concepts, by detecting and localizing GRBs with six 3U units. The successful realization of this experiment will guide the consolidation of the HFC design with the final goal to monitor the full sky and provide sub-arcmin localization of most GRBs.
The HERMES Pathfinder consist of two different projects: the HERMES Technological Pathfinder (HTP) and the HERMES Scientific Pathfinder (HSP). The former, funded by the Italian Ministry of University and Research (MIUR) and the Italian Space Agency (ASI), aims at producing three 3U CubeSats equipped with X/gamma-ray detectors [\citenum{Fuschino20,Evangelista20}]. The latter, funded by the European Union Horizon 2020 Research and Innovation Program, aims at realizing three additional units. Moreover, the project includes the design and development of the mission and science operation centers (MOC \& SOC). 
The HTP/HSP mini-constellation of six 3U units should provide enough GRB detections and localizations to:
\begin{itemize}
	\item validate the overall concept as well as to study the statistical and systematic uncertainty on both detection and localization to design the HFC;\\ 
	\item prove that accurate timing in the still poorly explored window $\mu$s-ms is feasible using detectors with relatively small collecting area;\\
	\item study uncertainties associated to the addition of the signal from different detectors to improve the statistics on high-resolution time series. 
\end{itemize}

Finally, ASI recently approved and funded the participation to the project SpIRIT (Space Industry Responsive Intelligent Thermal), founded by the Australian Space Agency, and led by University of Melbourne. SpIRIT will host an HERMES-like detector and S-band transmission systems. The HERMES-TP/SP mini-constellation of six CubeSats plus SpIRIT should be tested in orbit during 2022. More details on the overall HERMES mission description can be find in [\citenum{Fiore20}].

\begin{figure}
\begin{center}
\begin{tabular}{c}
\includegraphics[height=8cm]{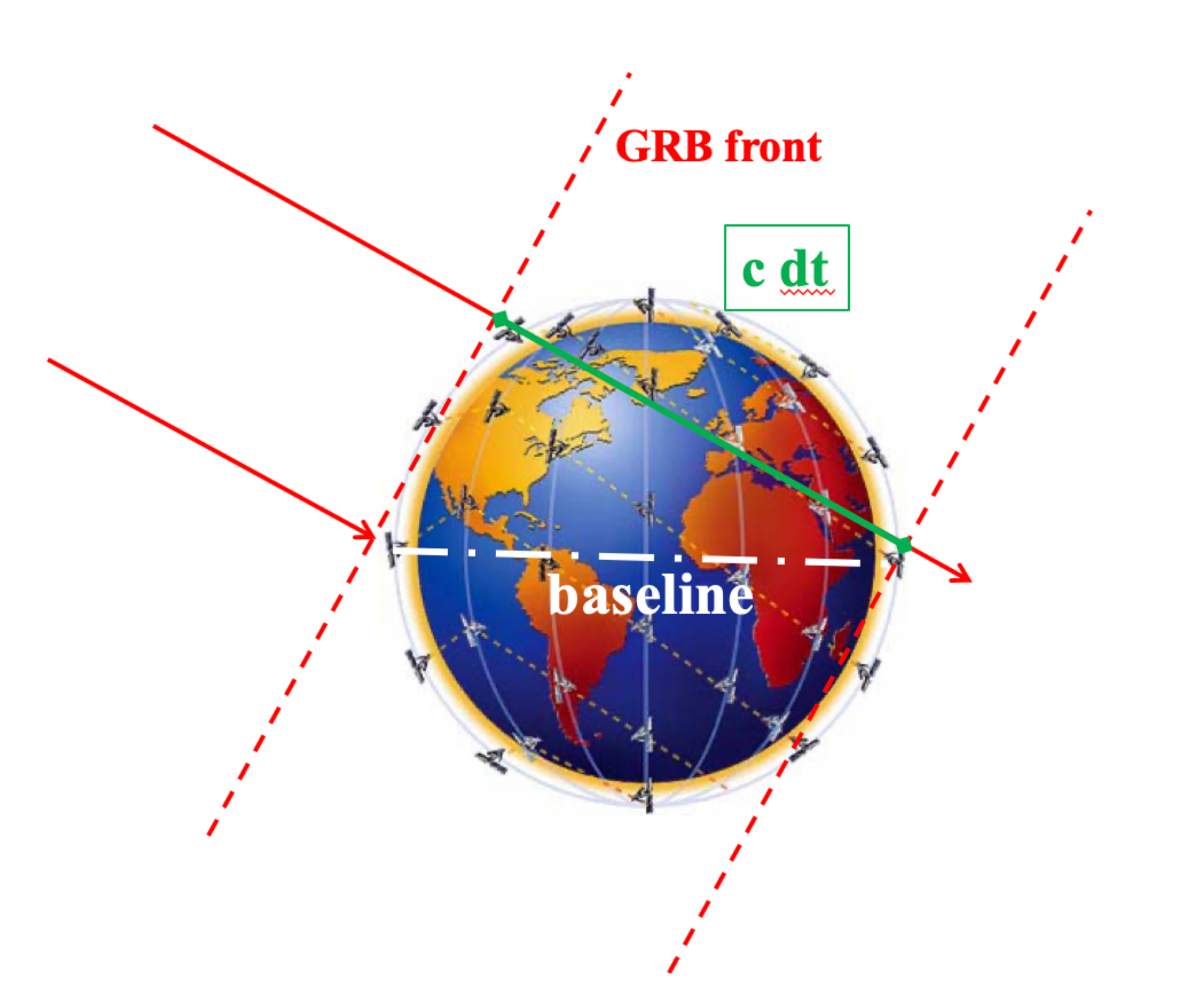} 
\end{tabular}
\end{center}
\caption[example] 
{ \label{fig:triangulation} 
Schematic representation of the \emph{Temporal Triangulation} principle applied to a constellation of CubeSats distributed in low Earth orbits. Red arrows represent the emitted GRB photons, while the red dashed lines describe the traveling GRB front wave at different times. The green segment represents the difference in travelled distances of the GRB photons detected by two CubeSats placed at a generic baseline (white dashed line).}
\end{figure}

\section{Triangulation technique}
The aim of this work is to investigate the localisation capabilities of the HTP/HSP mini-constellation composed of six 3U units. The main targets that will be discussed in the following are the highly energetic high-energy transients GRBs. The simple and yet robust idea that will be applied for accurately localise the transient astrophysical sources is the so-called \emph{Temporal Triangulation}.  

\begin{figure}
\begin{center}
\begin{tabular}{ccc}
\includegraphics[height=5cm]{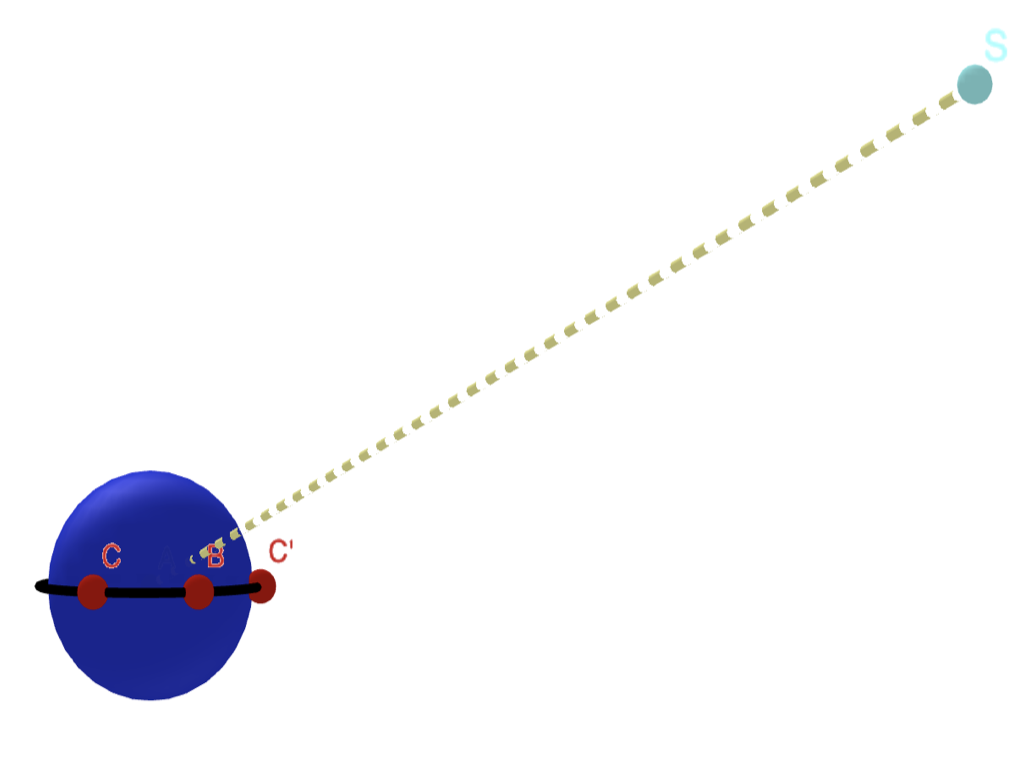} &
\includegraphics[height=6cm]{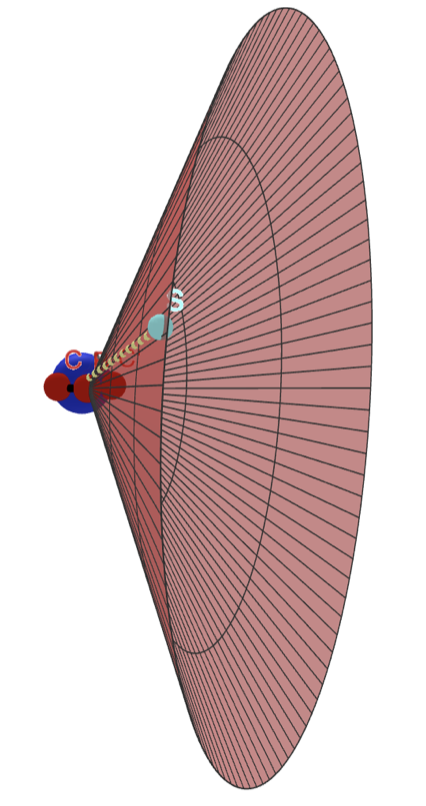} &
\includegraphics[height=6cm]{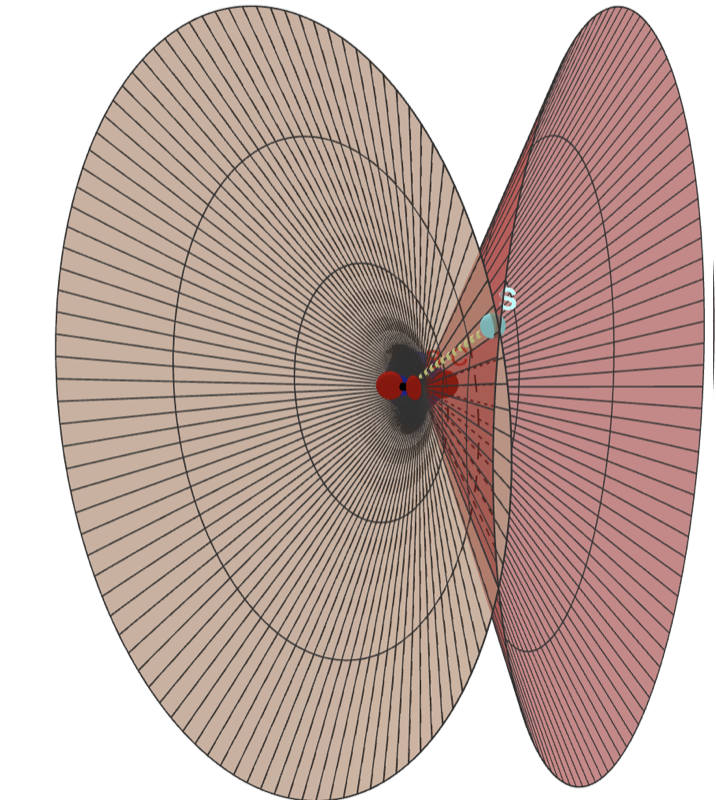} \\
a) & b) & c)\\
\end{tabular}
\end{center}
\caption[example] 
{ \label{fig:triangulation2} 
\emph{Panel a)} Schematic representation of a triplet of CubeSats distributed in an equatorial plane that observed the transient event located at the position $S$. \emph{Panel b)} Graphic representation of the source directions identified using the delay measured by combining the observations of the event obtained with the detectors B and C. \emph{Panel C)} Superposition of the source directions obtained combining the delays measured with the detector pairs B-C and B-C'. The intersection of the two cones identifies two possible locations of the event in the sky.}
\end{figure}

To describe the principle behind this method, let us represent the transient event as a narrow wavefront (pulse) traveling in a given direction and let us displace a network of detectors in space. The narrow wavefront will hit the detectors of the network at different times that depend on their spatial positions and the direction of the wavefront. As represented in Figure~\ref{fig:triangulation}, the transient event will be registered by two detectors of the network with a delay $dt$ that is proportional to their projected distance with respect to the source direction (green segment). The combination of the delays measured by different pairs of detectors observing the same event will allow to reconstruct its position in the sky. For the sake of description, let us consider a subset of 3 detectors distributed in the equatorial plane (panel \emph{a} of Figure~\ref{fig:triangulation2}) observing simultaneously an event $S$ located at a generic position in the sky (with direction with respect to the Earth barycenter represented by the yellow dashed line). As shown in panel \emph{b} of Figure~\ref{fig:triangulation2}, the delay measured $\Delta t_{BC}$ by combining the observations of the detectors B and C allows us to identify a set of infinite possible directions of the transient source forming a cone with circular base. Each of these source directions $\hat{\vec{d}}_i$ satisfies the relation $\Delta t_{BC}=\vec{\rho}_{BA}\cdot \hat{\vec{d}}_i / c$, where $\vec{\rho}_{BA}$ is the vector describing the distance between the detectors. A similar result can be obtained by combining the measurements of the detectors B and C. Interestingly, the superposition of these two results reduces the degeneracy on the source direction, allowing us to determine two possible positions of the source defined as the intersections of the two cones (Figure~\ref{fig:triangulation2} panel \emph{c}), one of which (as predicted) matches the source. By increasing the number of independent delay measurements, it is then possible to univocally localise the transient event in the sky. Therefore, it is clear that to achieve localisation capabilities, a generic fleet of detectors distributed in space should guarantee the simultaneous observation of an event with at least three of its elements.    

The description of the method reported so far does not take into account of any possible source of the uncertainty associated with the observational set-up. More in detail, the localisation capabilities of the system, hence the accuracy associated with the source position, will depend on several aspects such as the capability to reconstruct the position of the detectors during the observation of the event, the ability to recover delays between signals observed by different detectors and the capability of the detector to precisely time tag the photons associated with the transient event. 
A first proxy on the accuracy in determining the source position $\sigma_{PA}$ can be determined in the hypothesis of an event (e.g. a GRB) whose emitted photons arrive to a series of N detectors uniformly distributed in an orbit, and it is given by the expression:

\begin{equation}
\label{eq:sigma_pos}
\sigma_{PA} = \dfrac{\sqrt{(\sigma^2_{delay}+\sigma^2_{tpos}+\sigma^2_{time})}}{<Baseline>\sqrt{(n-1-2)}},
\end{equation}
where $\sigma_{delay}$ is the error on the delay measurement obtained combining the light-curves recorded by a pair of detectors, $\sigma_{tpos}=\sigma_{pos}/c$ is the error induced by the uncertainties on the spatial localisation of the detectors, $\sigma_{time}$ is the uncertainty on the absolute time reconstruction, $<Baseline>$ is the average distance between the detectors ad $n_{ind} = n-1$ is the number of statistically independent pairs of detectors used to determine the delay measurements.\\


For a more accurate approach on determining the position and relative uncertainties of a generic GRB in the sky by means of time delay measurements, let us again consider a swarm of $n$ satellites, each one identified by a position vector $\vec{r}_i$ (with $i=0,\dots, n-1$) with respect to a suitable reference frame, e.g. the Earth barycenter in equatorial coordinates.
To determine the GRB direction $\hat{\vec{d}}$, it is possible to measure the time delays of the GRB signals as seen from each pair of  satellites. 
Defining $t_0$ the time at which the GRB signal arrives at the origin of the chosen reference frame, each $i$-th satellite will receive the GRB at a time $t_i$
\begin{equation}
  t_i = t_0 - \frac{\vec{r}_i \cdot \hat{\vec{d}}}{c}.
\end{equation}
The expected time delays between two satellites will then be
\begin{equation}
    \Delta t_{ij}(\hat{\vec{d}}) \equiv t_j - t_i = \frac{(\vec{r}_j - \vec{r}_i) \cdot \hat{\vec{d}}}{c} =  \frac{\vec{\rho}_{ij} \cdot \hat{\vec{d}}}{c},
\label{eq:delayexp}
\end{equation}
where $\vec{\rho}_{ij} \equiv \vec{r}_j - \vec{r}_i$.
The real (measured) time delay between the signals recorded by two satellites $\Delta \tau_{ij}$ is inferred e.g. by applying cross-correlation techniques to the light-curves.
The direction of the GRB, e.g. its equatorial coordinates, can be estimated comparing the computed and measured delays between satellites using e.g. the non linear least squared method.
We define the $\chi^2(\hat{\vec{d}})$ function as the sum of the squares of the differences between the expected and observed time delays divided by its statistical error
  \begin{equation}
    \chi^2(\hat{\vec{d}}) = \sum_{i = 0}^{n - 2}\sum_{j = i + 1}^{n - 1} \frac{(\Delta \tau_{ij} - \Delta t_{ij}(\hat{\vec{d}}))^2}{\Theta_{ij}^2},
    \label{eq:chisq}
  \end{equation}
where $\Theta_{ij}^2$ includes the positional error on the satellites expressed in light-seconds, the accuracy in the absolute timing of the detectors, the uncertainty on constraining the time delay between the signals and any hypothetical systematic uncertainty related to the set-up or method applied.
  
The unitary vector $\hat{\vec{d}}$ identifying the GRB direction can be then written in terms of the equatorial coordinates Right Ascension $\alpha$ and Declination $\delta$, that is
  
  \begin{equation}
    \hat{\vec{d}} = \{\cos{\alpha} \cos{\delta}, \sin{\alpha} \cos{\delta}, \sin{\delta}\}.
  \end{equation}

Minimizing Eq.~\ref{eq:chisq} with respect to $\alpha$ and $\delta$ gives us an estimate of the direction of the GRB.
Moreover, if Eq.~\ref{eq:chisq} satisfies all the hypotheses in [\citenum{Avni76}], we can also calculate the confidence region for the GRB equatorial coordinates on the plane of the sky.

\section{EXPLOITING THE HERMES PATHFINDER LOCALISATION CAPABILITIES }

In the following, we describe in details the analysis performed, as well as the related assumptions and caveats, to investigate the capabilities of the HTP/HSP mini-constellation six 3U CubeSats on localizing GRBs in the sky.

The key to accurately locate an event by means of the temporal triangulation method described above is to decrease as much as possible the uncertainties summarized by the term $\Theta^2$ in Eq.~\ref{eq:chisq}. Starting from the HTP/SP technical properties, we can investigate the positional uncertainty budget to be able to identify the most crucial limiting factors. The final design of the spacecraft including GPS receivers and accelerometers, guarantees the possibility to reconstruct the position of the CubeSats with an accuracy smaller than 30 meters, that translates into a temporal accuracy lower than 30 ns [\citenum{Fiore20}]. Moreover, the absolute timing accuracy achievable from the detector is going to be lower than 0.4 $\mu$s for both the X and S modes [\citenum{Evangelista20}]. As we will discuss in more detail later, considering the HTP/HSP set-up, we can conclude that, even for the brightest GRBs, the uncertainty on the GRB position will be dominated by the accuracy on the time delay measurement between the GRB light-curves and possibly by unknown systematics still to be investigated. In the following we will show that on average uncertainties on the time delays obtained applying cross-correlation techniques are of the order of tenths of milliseconds, a few orders of magnitude larger than the other uncertainties discussed above.

\subsection{GRB structure and time delay accuracy}
\label{sec:grbcc}
To be able to investigate the accuracy on the measurement of the time delays between the arrival times for photons emitted by a generic GRB and observed by different detectors of the HTP/HSP mini-constellation, we built a procedure that includes the creation of GRB templates, the application of cross-correlation techniques, as well as Monte Carlo simulations.

As a first step, we searched the available Fermi GBM archive seeking for GRBs characterized by variability on time scales as short as a few milliseconds. The hypothesis being that, fast variability should enhance the sensitivity on time delay measurements, especially when the statistics of the available data is relatively limited. We then isolated two candidates, one belonging the so-called short GRBs and the other from the long GRBs. More specifically, the short GRB (id. GRB120323507) has been observed on 2012 March 23, and it is characterized by a $t_{90}$\footnote{Time interval in which the integrated photon counts increase from 5\% to 95\% of the total counts.} duration of $\sim0.4$ seconds with a fluence of $\sim1\times10^{-5}$ erg cm$^{-2}$. On the other hand, the long GRB (id. GRB130502327) has been detected on 2013 May 2, and it is characterized by a $t_{90}$ duration of $\sim24$ seconds with a fluence of $\sim1\times10^{-2}$ erg cm$^{-2}$.

The data collected from the GBM catalogue includes light-curves from the sodium iodide (NaI) scintillators and from the cylindrical bismuth germanate (BGO) scintillators, both having a collecting area of about 125 cm$^2$ [\citenum{Meegan09}]. The NaI detectors are sensitive to energies included between few keV up to about 1 MeV, while the BGO detectors cover the energy range from 150 keV to 30 MeV. We selected data captured with the so called \textit{Time-tagged event (TTE)} format, where the GRBs are continuously recorded with a time resolution of 2 \us, within a time interval that includes 15-30 s of pre-trigger information and about 300 s of data after the trigger time.

To be able to recreate the GRB light-curves as seen by the HTP/HSP detectors, we selected the energy range 50-300 keV at which corresponds to the largest effective area of scintillators (around 50 cm$^2$; see [\citenum{Campana20}] for more details). For this reason, data collected from the GBM observations has been previously filtered in order to retain events only in this energy range.

The need to generate GRB templates (functional forms of the GRB light-curves) comes from two crucial aspects in the procedure followed to investigate the measurement of signal delays: a) flexibility to recreate GRB light-curves independently of the detectors effective area and b) the possibility to simulate GRB light-curves with intrinsically poor statistics.  

\begin{figure*}[h!]
\centering
\includegraphics[scale=0.52,angle=0]{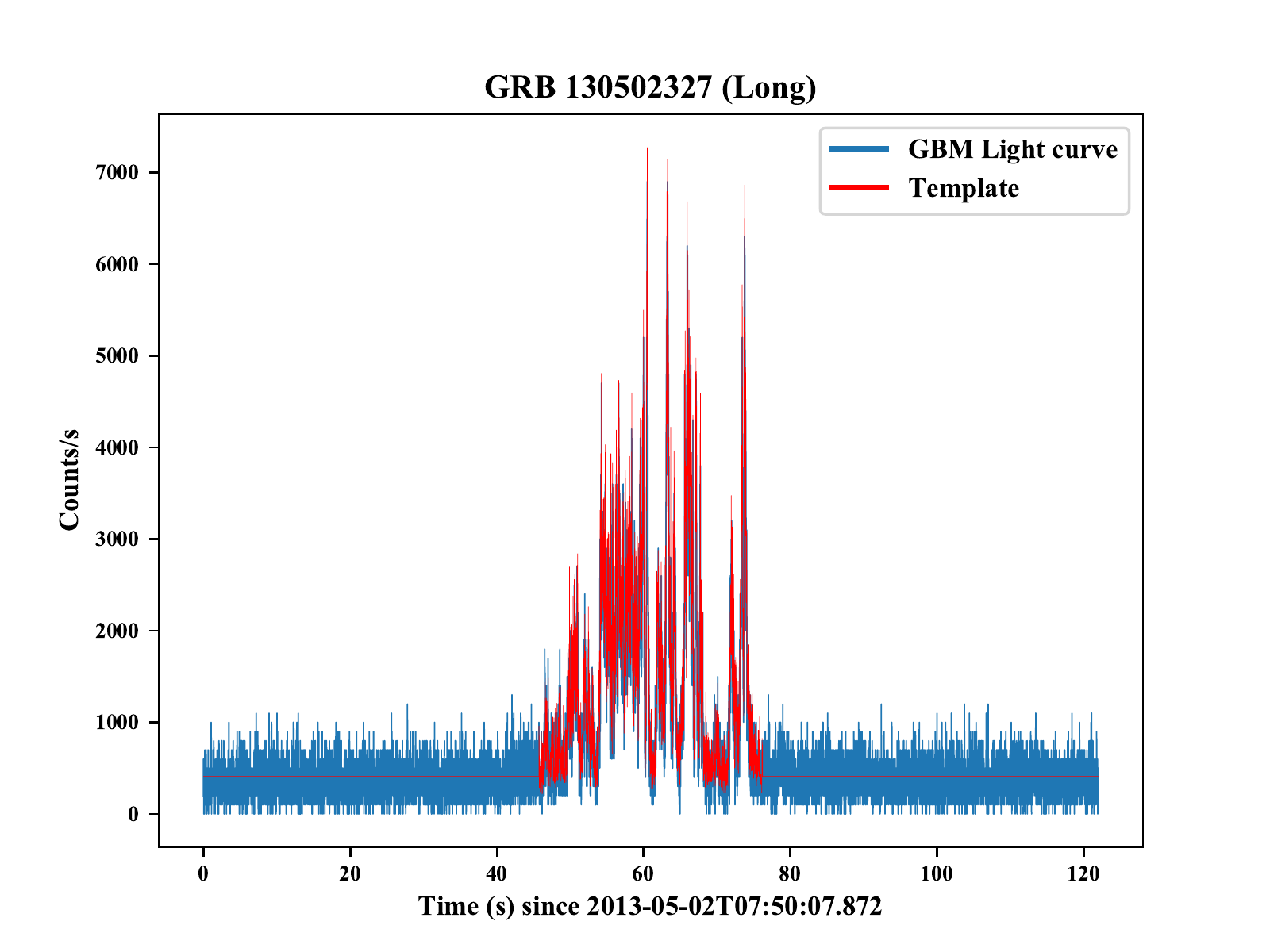}
\includegraphics[scale=0.52,angle=0]{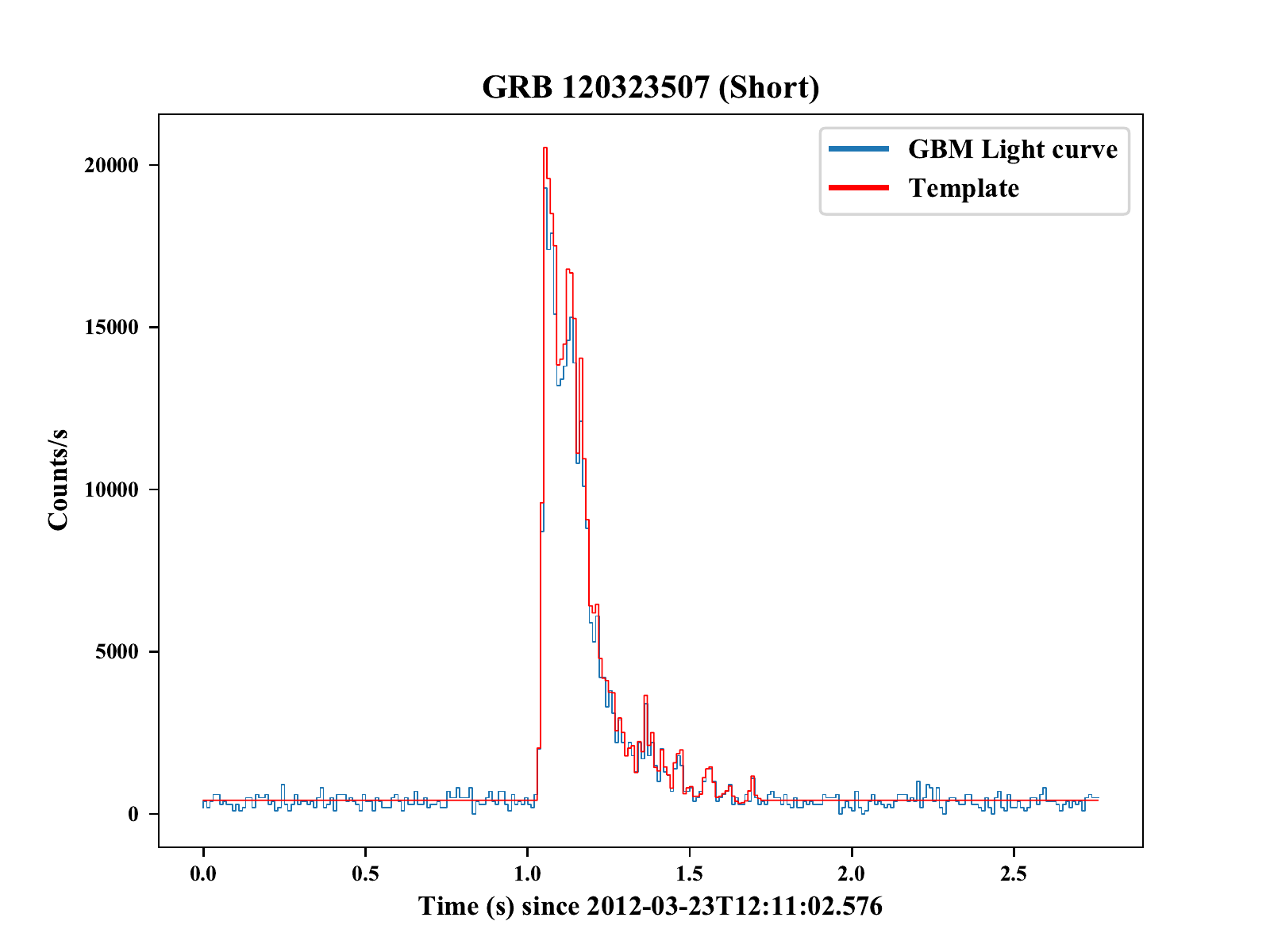}

\caption{The Fermi GBM light-curves of the long GRB 130502327 (left panel) and of the short GRB 120323507 (right panel) and the relative template (red line) obtained with the procedure described in the text. In both cases, for reasons of clarity, we applied a bin time of 10$^{-2}$ s for the light-curves and templates on both the GRBs.} 
\label{fig:template_short_long}
\end{figure*}

Indeed, simulations on short time scales ($\sim$1 ms) of a unique-like type of transient events (such as a GRB) based on observed light-curves, can be challenging when the effective area of the detector is so small that the statistic is fully dominated by Poissonian fluctuations that unavoidably characterize the detection process. In particular, if the detected counts within the given time scale is $\leq1$, fluctuations of the order of 100\% are expected. If, naively, the number of counts per bin is simply rescaled to account for an increased effective area, these fluctuations can introduce a false imprint of 100\% variability with respect to the original signal. No definite cure is available to mitigate this problem, that could be, however, alleviated by binning and/or smoothing techniques. Although smoothing techniques allow the creation of light-curves for a desired temporal resolution, correlation between subsequent bins is unavoidable. Cross-correlation techniques are strongly biased by this effect, therefore, we opted for a more conservative method implying standard binning in which the number of photons accumulated in each (variable) bin is fixed. After several trials and Monte Carlo simulations, we found that 6 photons per bin allows to preserve the signal variability introducing undesired fluctuations not larger than $\sim$30\%. Applying this binning technique to the GBM light-curves (at the maximum time resolution of 2 \us), we generated variable bin size light-curves. In order to generate a template usable on any timescale, we linearly interpolated the previous light-curve to create a functional expression (template) for the theoretical light-curves. We note explicitly, that linear interpolation between subsequent bins is the most conservative approach that does not introduce spurious variability on any timescales. For a given temporal bin size, it is then possible to rescale the GRB template previously described in order to match the requested effective area (e.g. that of the HTP/HSP detectors), generating then the expected number of photons within the time bins. In addition, before and after the burst, we rescaled the background on the GRB template to match the nominal background collected by the detector estimated by studying the selected CubeSat orbit [\citenum{Campana20}]. Figure~\ref{fig:template_short_long} shows the templates (red lines) for the long (left panel) and short (right) GRBs generated by following the procedure described above.

\begin{figure}[h!]
\centering
\includegraphics[scale=0.52,angle=0]{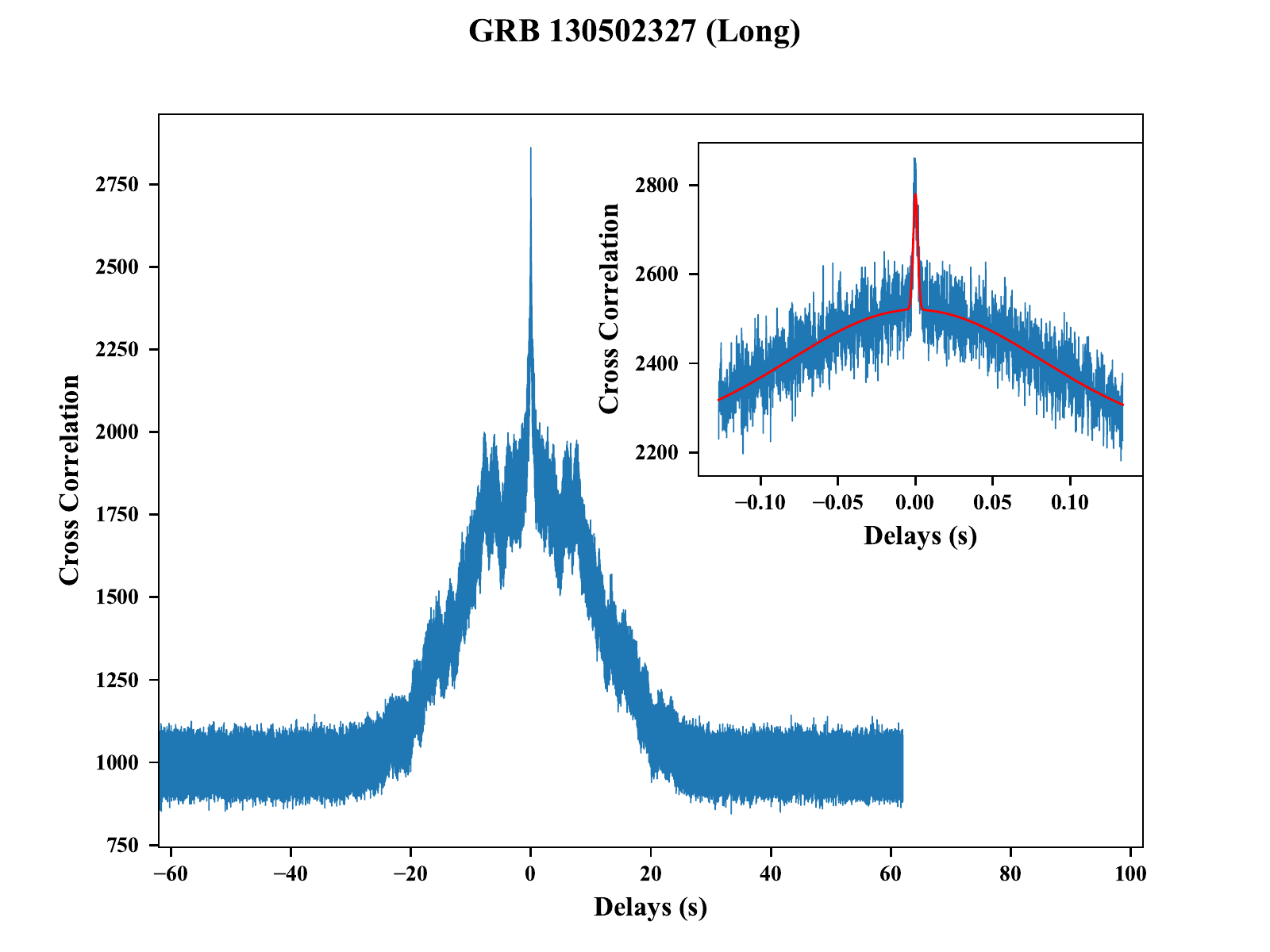}
\includegraphics[scale=0.52,angle=0]{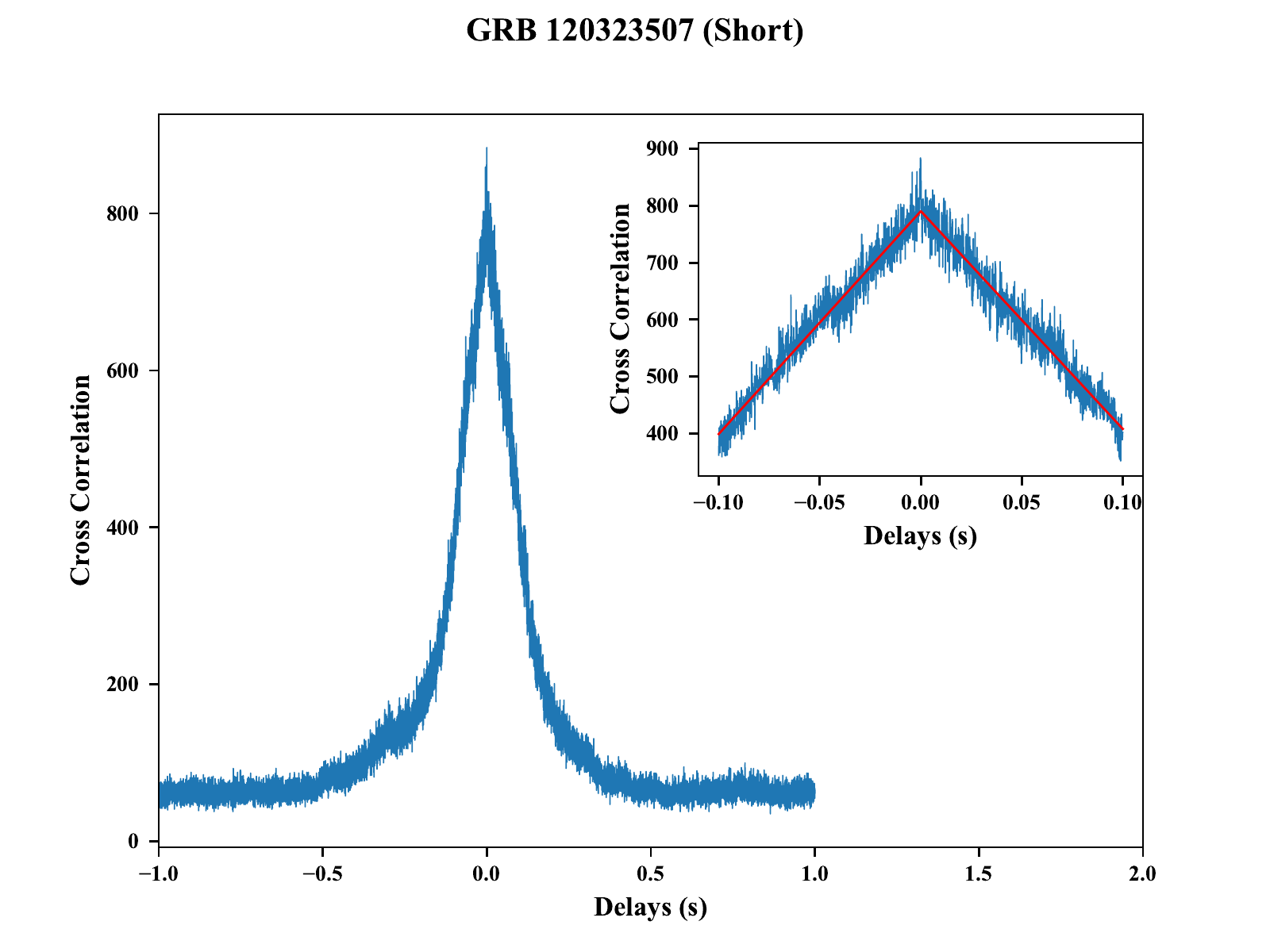}

\caption{Cross-correlation functions obtained by simulating the GRB light-curves of the long GRB 130502327 (left panel) and of the short GRB 120323507 (right panel) using the templates shown in \autoref{fig:template_short_long} rescaled to match the effective area of the HTP/HSP detectors. The insets report a zoom-in of the cross-correlation profiles around the peak as well as their best-fitting model (red solid line).} 
\label{fig:xc_profiles}
\end{figure}

Starting from the GRB templates, we generated light-curves by rescaling the detector effective area to match that of the HTP/HSP detectors and by applying a Poissonian randomization of the counts contained in each bin of the template. 
We then applied standard cross-correlation techniques (see e.g. [\citenum{bendat11,ianniello82}]) on two light-curves with the aim to determine the time delay between the signals. Since we are interested in reconstructing the accuracy achievable in the time delay, and not strictly on the delay value per se, we did not shift in time the template to simulate the signals. 
To extract the temporal information of the delay, we then fitted a restricted region around the peak of the cross-correlation function with an \emph{ad hoc} model. 
The uncertainty associated with the location (in delay) of peak of the cross-correlation function defines the accuracy on determining the delay between the two detected signals. 
In Figure~\ref{fig:xc_profiles}, we report the cross-correlation functions obtained with the simulated light-curves of the long GRB 130502327 (left panel) and the short GRB 120323507 (right panel) as seen by the HTP/HSP detectors and assuming an on-axis detection. Moreover, in the insets of Figure~\ref{fig:xc_profiles} we report a zoom-in of the peak of the cross-correlation functions and its relative best-fitting model (red solid line).

\begin{figure}[h!]
\centering
\includegraphics[scale=0.45,angle=0]{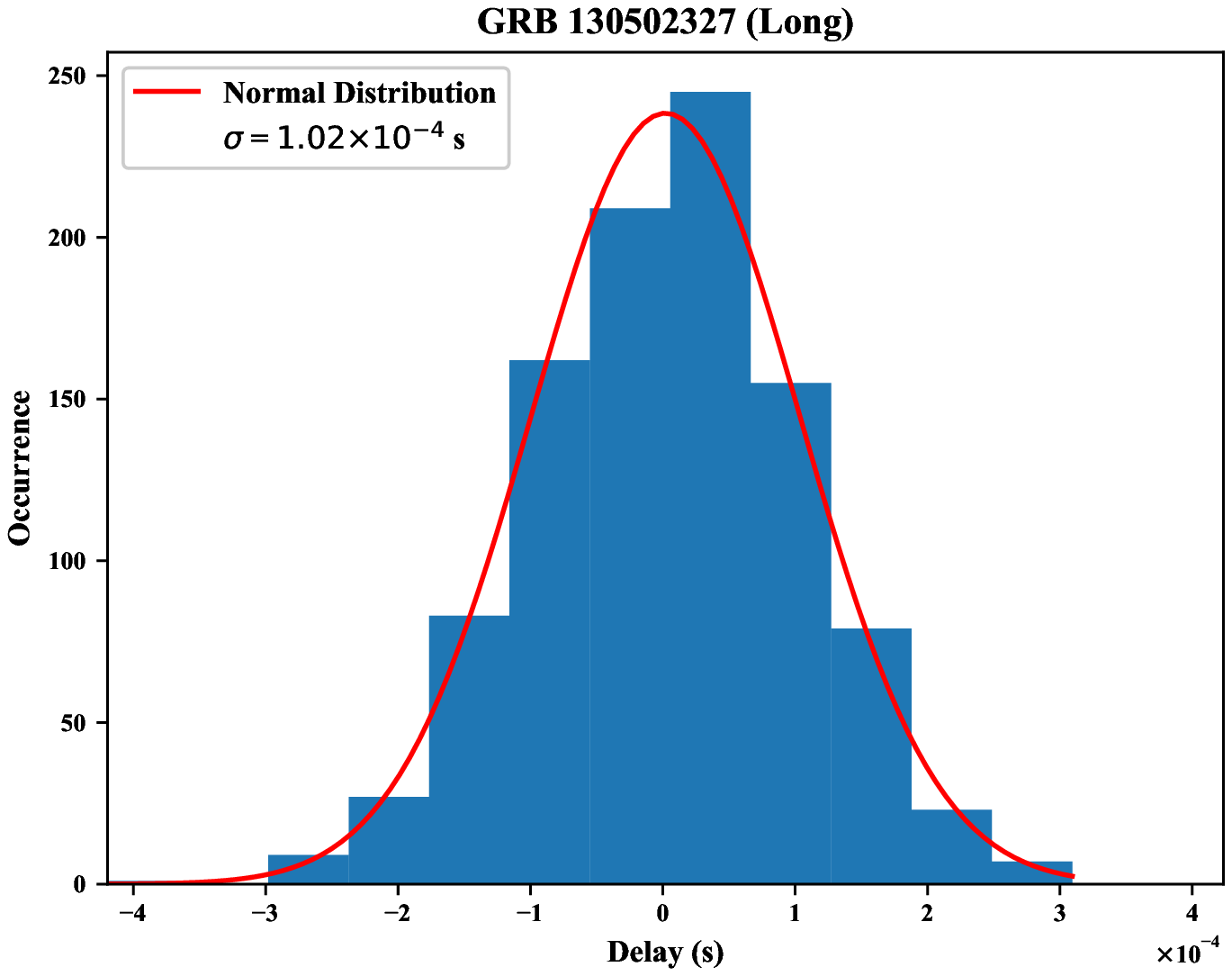}
\includegraphics[scale=0.45,angle=0]{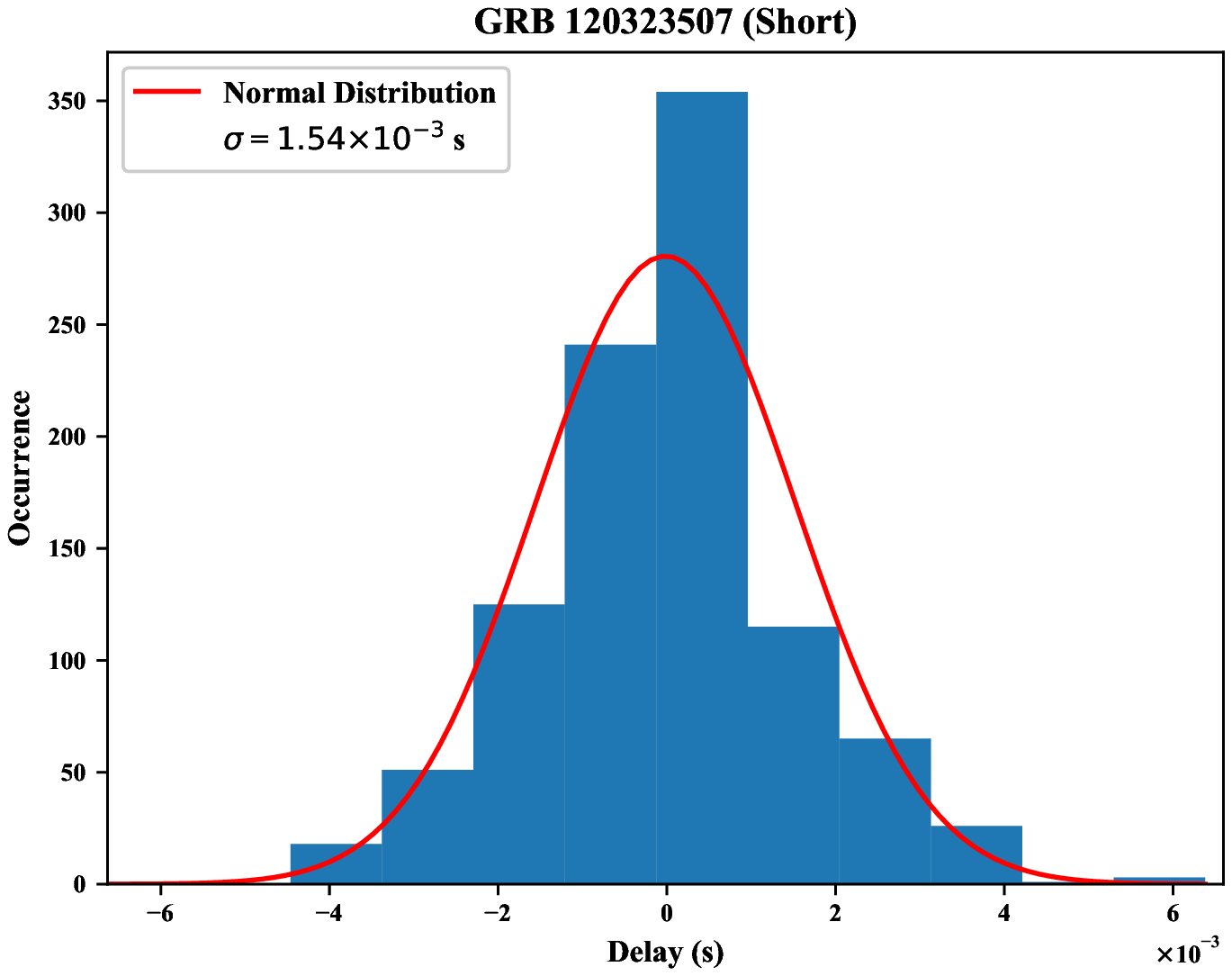}
\caption{Distribution of delays obtained applying cross-correlation techniques to pairs of simulated light-curves of the long (left panel) and short (right panel) GRBs rescaled to match the HTP/HSP effective area. The distributions summarize the result of the 1000 Monte Carlo simulations (see text for more details). The overlaid red line represents the best-fitting normal distribution to the data.} 
\label{fig:xc_sigma}
\end{figure}

\emph{How reliable is the fitting of the cross-correlation peak in terms of determining the accuracy of time delays between the two light-curves?} Given the complexity of an analytic approach to the problem, we decided to tackle the issue taking advantage of Monte Carlo simulations. More precisely, for each GRB, we generated 2000 light-curves by means of Poissonian randomization of the template. For each of the 1000 cross-correlation functions generated, we then determined the delay by fitting its peak as described above. From the overall distributions of delays obtained for the long and short GRBs (left and right panels in \ref{fig:xc_sigma}), we estimated the standard deviations $\sigma_{cc-short}\sim1.54\times10^{-3}$s and $\sigma_{cc-long}\sim 1.02\times10^{-4}$s, respectively, that we interpreted as a realistic estimate of the accuracy on the time delay measured with this procedure. It is worth noting that for the long GRB the mean uncertainty (obtained by averaging the results from the 1000 simulations) on the time delay obtained by fitting the peak of the cross-correlation function is only 20\% smaller with respect to the sigma of the time delay distribution. On the other hand, for the short GRB the uncertainty on the fit of cross-correlation peak is almost a factor of 4 smaller with respect to the sigma of the peak distribution.

\emph{How does the GRB morphology affect the capability to accurately determine time delays by applying cross-correlation techniques?}
As a first attempt to test the dependence of the cross-correlation uncertainty on the brightness and temporal structure of the GRBs, we applied the technique described above to two unbiased random samples each including 100 short and 100 long GRBs selected from the available Fermi GBM catalogue.
The randomness of the samples guarantees a good coverage of the vast variety of phenomenologies, fluxes, durations and intrinsic variability that were recorded during the Fermi mission up to the moment in which this paper was written. For the long GRBs, the sample includes bursts having background subtracted fluxes ranging between 0.16 and 26 ph cm$^{-2}$ s$^{-1}$, durations between 3 and 138 s, and fluence values between 3.2 and 638 ph cm$^{-2}$. On the other hands, the sample of short GRBs ranges in flux between 0.6 and 188 ph cm$^{-2}$ s$^{-1}$, durations between 0.03 and 1.9 s and fluence values between 0.2 and 75 ph cm$^{-2}$.\\

\begin{figure}
\begin{center}
\begin{tabular}{cc}
\includegraphics[height=6cm]{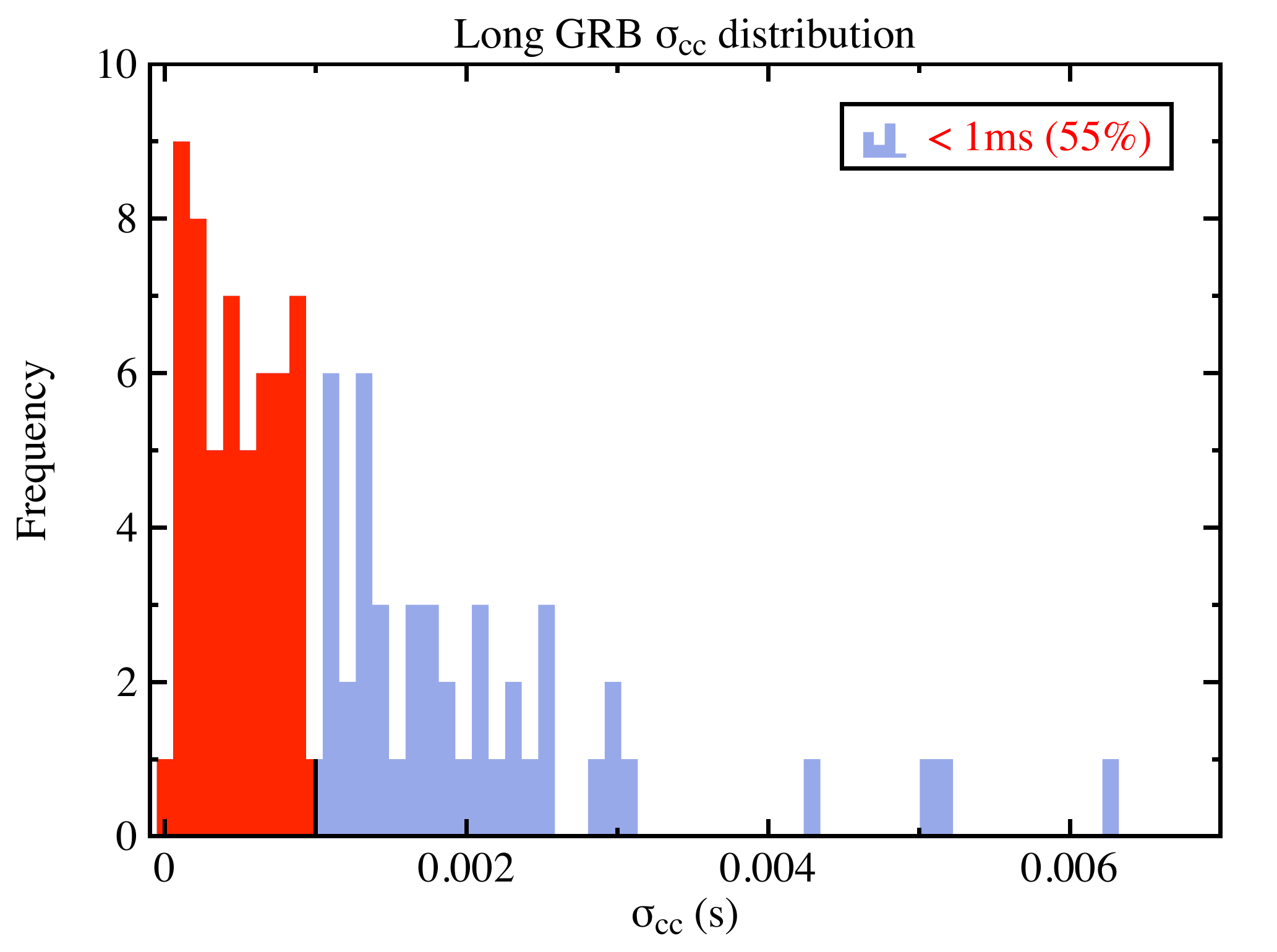} &
\includegraphics[height=6cm]{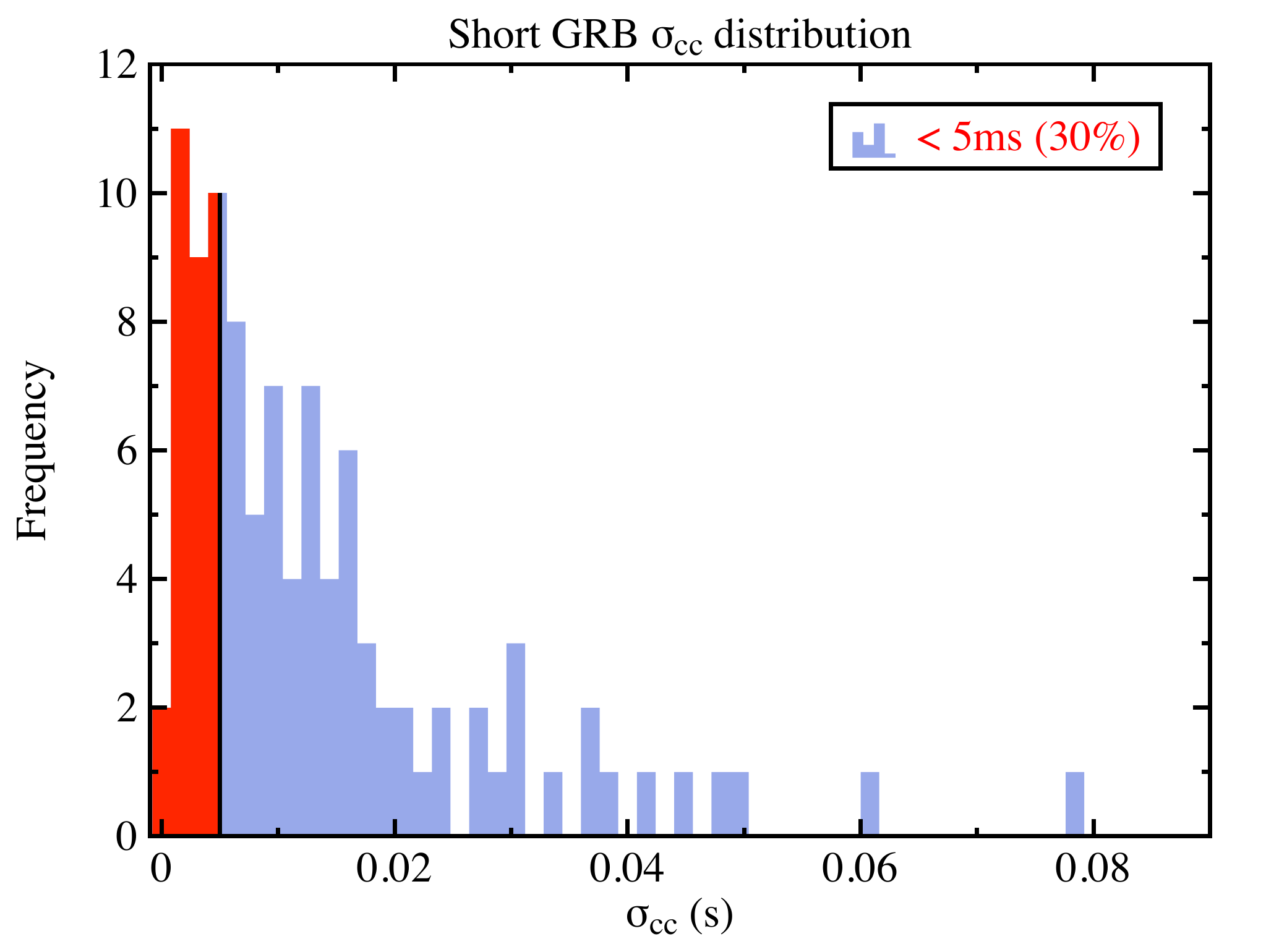} \\
\includegraphics[height=6cm]{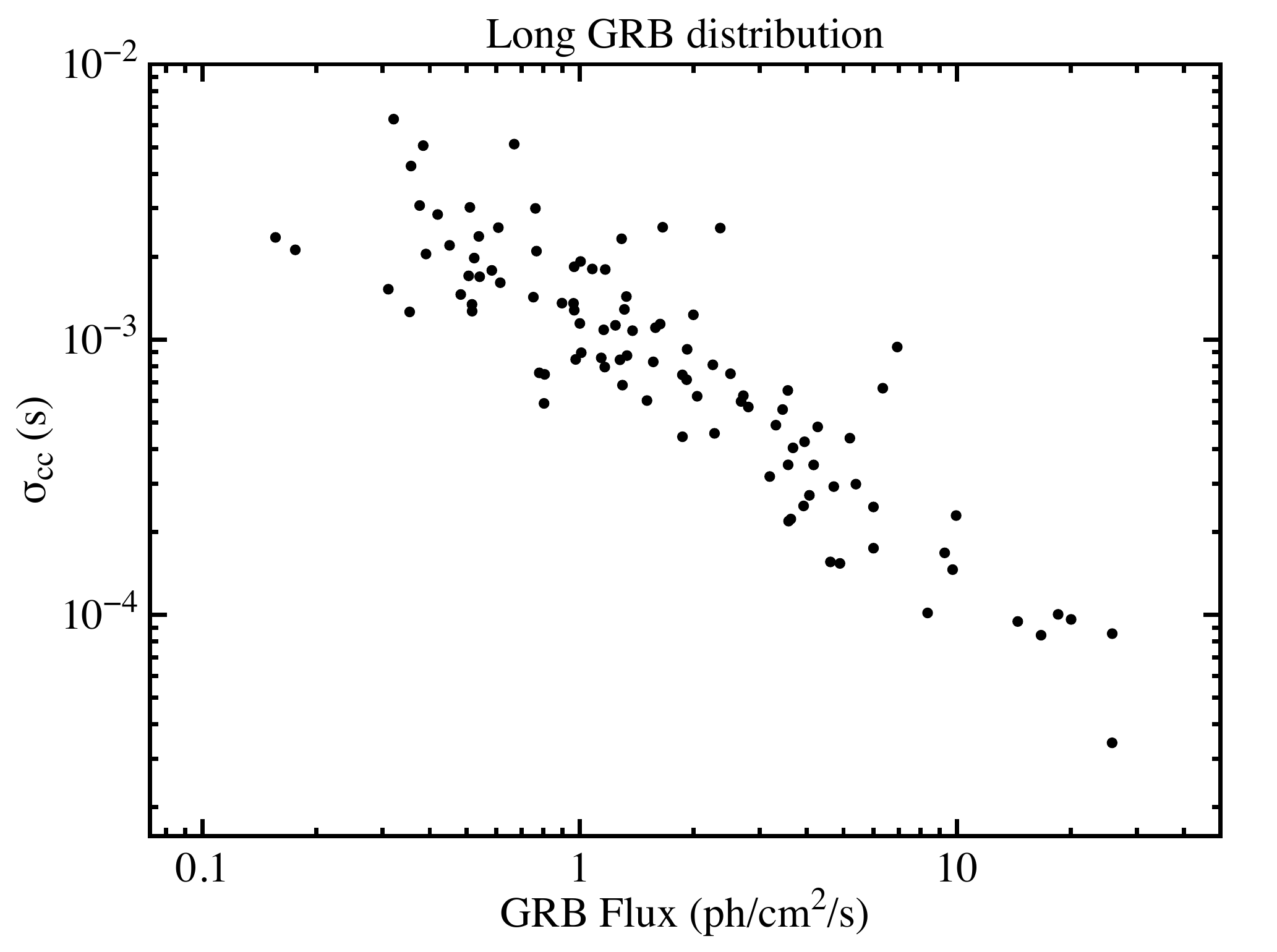} &
\includegraphics[height=6cm]{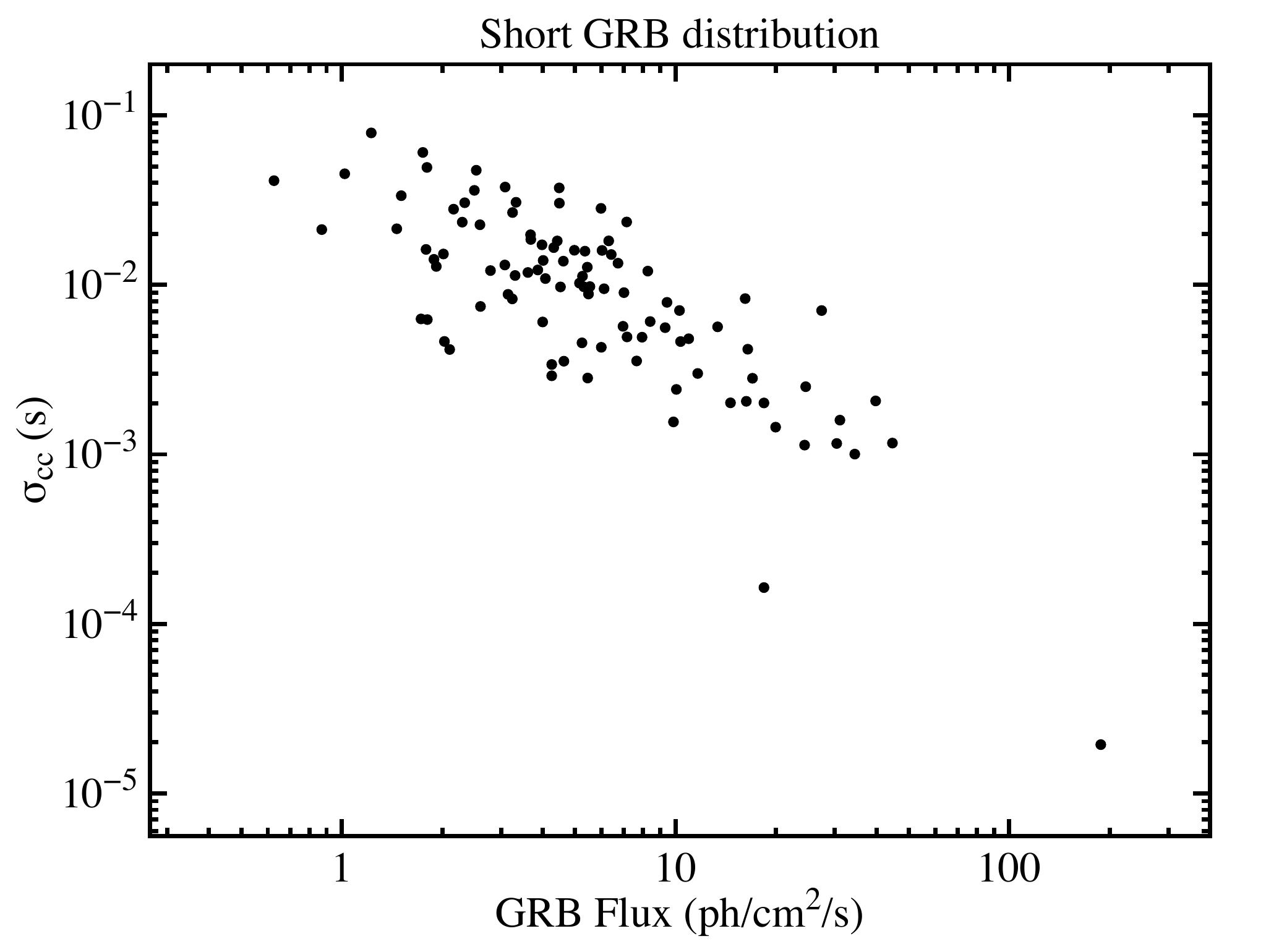} \\
\end{tabular}
\end{center}
\caption[example] 
{ \label{fig:sigma_sample} 
\emph{Top Left panel:} distribution of the delay accuracy estimated via cross-correlation techniques of a random sample of 100 long GRB selected from the Fermi GBM catalogue. In red we highlighted the sub-sample (55\%) characterized by $\sigma_{cc} \leq 1$ ms. \emph{Top Right panel:} distribution of the delay accuracy estimated via cross-correlation techniques of a random sample of 100 short GRB selected from the Fermi GBM catalogue. In red we highlighted the sub-sample (30\%) characterized by $\sigma_{cc} \leq 5$ ms. \emph{Bottom Left panel:} cross-correlation accuracy as a function of the GRB fluxes for the sample of long GRB. \emph{Bottom Right panel:} cross-correlation accuracy as a function of the GRB fluxex for the sample of short GRB.}
\end{figure}

For each burst, we simulated 2000 light-curves that allowed us to generate 1000 cross-correlation functions. Following the procedure described above, we then determined the time delay by fitting the interval near the peak with an \emph{ad hoc} model function (e.g. Gaussian functions, combination of two Gaussian profiles having a common centroid and asymmetric double exponential functions). The top left (top right) panel in Figure~\ref{fig:sigma_sample} shows the distributions of the time delay uncertainties (each representing the standard deviation of 1000 Monte-Carlo simulations) estimated cross-correlating the sample of long (short) GRBs. We note that an accuracy equal or smaller than 1 ms is obtained for 55\% of the long GRBs investigated (Figure~\ref{fig:sigma_sample} top-left panel, red area), while an uncertainty equal or smaller than 5ms is obtained for 30\% of the short GRBs (Figure~\ref{fig:sigma_sample} top-right panel, red area). Finally, the bottom left and right panels of Figure~\ref{fig:sigma_sample} represent the dependence of the cross-correlation accuracy as a function of the GRB flux for the long and short GRB, respectively. As expected, stronger GRBs allow to recover time delays with a better accuracy.

\subsection{Mission scenario}
Testing the localisation capabilities of the HTP/HSP mini-constellation requires the definition of a specific mission scenario. In the following, we will give a brief description of the experimental set-up used to the analysis described in this work, that represents the outcome of a detailed mission analysis investigation performed during the first year of the project (see [\citenum{Colagrossi20}] for a detailed description).

\subsubsection{Low Earth Equatorial Orbit}
To achieve to scientific goals of the mission, the HTP/HSP orbit has been accurately investigated and finally restricted to a low Earth orbit with altitude ranging between 500 and 600 km and inclination $<20^\circ$. Within the specific framework of the analysis reported here, the adopted reference orbit has the following properties:

\begin{itemize}
\item altitude h=550 km;
\item circular orbit (eccentricity = 0);
\item equatorial orbit (inclination = 0).
\end{itemize}

\subsubsection{Space segment injection strategy}
The space segments injection strategies explored in the mission analysis are the following:
\begin{itemize}
\item Dedicated multiple injections - one per satellite - into different true anomalies at $t_0$; 
\item Single injection of each triplet with imposed relative motion among the spacecraft belonging to the same triplet, imposed by the deployer release spring.
\end{itemize} 

As resulted from the analysis, the first option is highly sensitive to the release conditions, e.g. natural perturbations cause a relative drift which is emphasized by the launcher and deployer injection uncertainties jeopardizing this strategy robustness in terms of scientific outcome. This option requires a dedicated launch for the HTP/HSP mission. 
On the other hand, the second option is feasible without a dedicated launch, since the triplet elements can be released in a single launch event with no dedicated injection maneuver. The relative motion between the satellites is actually imposed by the deployer spring authority which overcomes the natural perturbations effects and the launcher injection uncertainties, making more reliable the expected scientific outcomes foreseen in the design phase. 
Here, we will consider the second injection strategy to perform the localization test.

\begin{figure}[h!]
\centering
\includegraphics[scale=0.45,angle=-90]{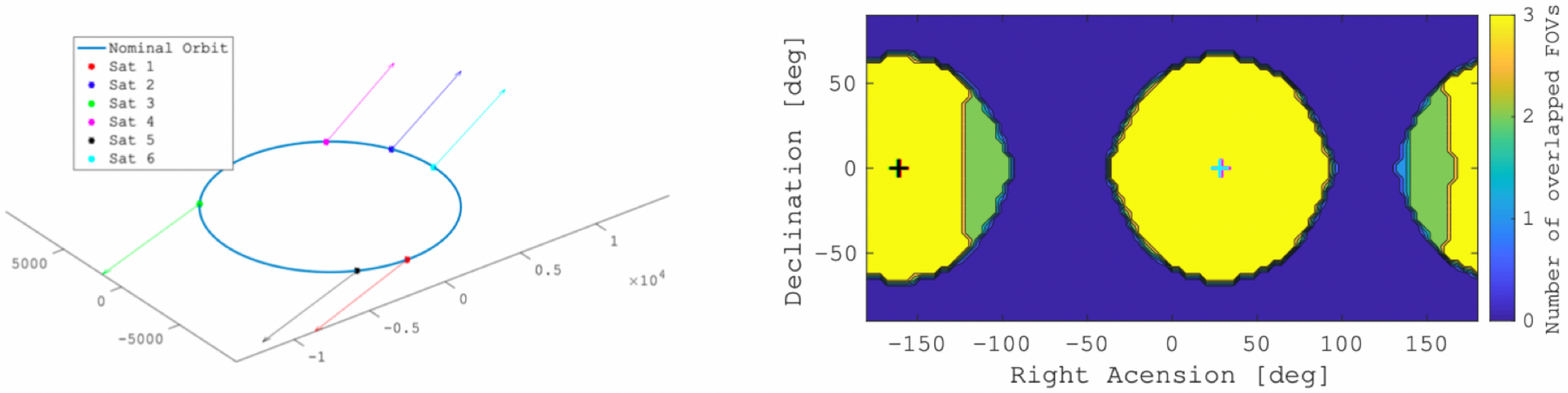}
\vspace{-2.5cm}
\caption{LVLH optimal pointing strategy} 
\label{fig:strategy}
\end{figure}

\subsubsection{Pointing strategy}
\label{sec:po_strat}
The selection of the nominal pointing strategy significantly affects the HTP/HSP performances in terms of number of detected and localized GRBs. During the study of the mission analysis and the spacecraft design activity, specific trade-off studies have been carried out on the topic, taking advantage of the quite enhanced attitude control performances of the HERMES CubeSats. In fact, the pointing direction of the payloads Line Of Sight (LOS) can be controlled and even varied along the mission time-line, according to the short-medium planning for the payload utilization, compliant with the space segment capabilities, to maximize the science mission outcome. 
Three different pointing strategies have been explored:
\begin{itemize} 
\item Zenith pointing for each payload LOS; 
\item Co-alignment of $n\geq3$ payload LOSs on an Inertial-selected direction; 
\item Co-alignment of $n\geq3$ payload LOSs on a LVLH-selected direction (i.e. LOSs aligned on the zenith direction of a specific satellite in the fleet).
\end{itemize} 

The third pointing strategy, that will be used for the analysis discussed in this document, is preferred to maximize the scientific outcome of the mission. To do that, periodical optimization of the LOS direction of each satellite are foreseen in order to maximize the overlapping Field of View (FoV) and hence the number of GRBs potentially triangulated. More in details, the whole mission is divided in periods (from days to weeks) in which the pointing directions of each satellite is kept fixed in the non-inertial LVLH (Local Vertical-Local Horizontal) reference frame. The pointing direction of each satellite in the LVLH frame is uniquely defined by two angles: the first, in the orbital plane, is the angular displacement between the LOS and the radial direction, and the second is the elevation of the LOS above (or below) the orbital plane. During each period, the optimal set of angles (two for each satellites) is selected using a heuristic particle-swarm optimization algorithm to maximize the scientific return, i.e. the number of GRBs triangulated during that frame time. Figure~\ref{fig:strategy} shows the position and pointing of the HTP/HSP detectors within the LVLH pointing strategy (left panel) as well as the associated map of the overlapping FoV.

\subsubsection{Simulation strategy}
\label{sec:strat}
To investigate the level of accuracy on the detection and localization of GRBs with respect to the mission scenarios previously described we adopted the following strategy: 
\begin{itemize}
	
\item based on the results reported in the Fourth Fermi GBM GRB catalogue [\citenum{vonKienlin20}], we generated GRB events assuming an uniform distribution in the plane-of-the-sky. The number of simulated GRBs during the mission lifetime (2 years) reflects the Fermi GBM detection rate $\alpha_{GBM} \simeq 0.083$ GRB/sr/d, which corresponds approximately to a total of 760 GRB events in the whole sky ($4\pi$ steradians) within the assumed time interval; 

\item we temporally located the 760 simulated GRB events by uniformly sampling at random from the time data-set associated to the positions of the fleet elements. More specifically, we randomly extracted 760 time-intervals (1-minute length) among the 1051201 available from the simulation of the satellite positions. With that, we are assuming a GRB detection process with duration shorter or equal to 1 minute. We note that this assumption is surely valid for short GRBs, whilst it is not applicable for 30\%-40\% of the long GRBs showing duration larger than 60 seconds; 

\item for each GRB event, we determined the number of active satellites (non-transiting within the South Atlantic Anomaly) able to detect it, by verifying that the direction of the GRB and the LoS of the instrument identify an angle lower or equal $60^\circ$. This guarantees that the GRB event falls within the $3\pi$ steradians FoV (Full Width at Half Maximum) of the detector. 

\item we processed only GRB events observed by a minimum of 3 satellites, or equivalently, for which a minimum of 3 satellites have the GRB in their FoVs. This guarantees the possibility to apply temporal triangulation techniques to determine the position of the GRB in the sky. 

\end{itemize}

For each of the GRBs potentially localizable, we determined the expected time delays $\Delta t_{ij}(\hat{d})$ for the independent pairs of satellites by using Eq.~\ref{eq:delayexp}. The associated real (measured) time delay $\Delta \tau_{ij}$ will be then generated as:
\begin{equation} 
     \Delta \tau_{ij}=\Delta t_{ij}(\hat{\vec{d}})+N(0,\sigma_{cc}),
\end{equation}              
where $N(0,\sigma_{cc})$ is the cross-correlation uncertainty randomly extracted assuming a normal distribution with zero mean and standard deviation equal to the results obtained from the analysis of the sample of long and short GRBs described above. More specifically, among the 760 GRBs, 83\% (following to Fermi GBM statistics) were simulated assigning the long GRBs accuracy $\sigma_{cc-long} = 1$ ms, while the remaining 17\% will be simulated as short GRBs with $\sigma_{cc-short} = 5$ ms.
Using Eq.~\ref{eq:chisq}, we then calculated a $\chi^2$ map of the whole sky by creating a grid of points uniformly sampling the equatorial coordinates. Finally, we estimated the most probable position of the simulated GRB by calculating the values of $\alpha$ and $\delta$ that minimize the $\chi^2$ function. Confidence intervals $\sigma_\alpha$ and $\sigma_\delta$ were estimated at $1\sigma$ level considering a $\Delta \chi^2 (\hat{d}) = 
\Delta 
\chi^2(\nu, 68\%)$, where $\nu$ is the number of parameters in the function. We emphasize that the coordinate intervals obtained by marginalizing the confidence region might be overestimated, especially $\sigma_\alpha$.

\begin{figure}
\begin{center}
\begin{tabular}{cc}
\includegraphics[scale=0.4,angle=0.0]{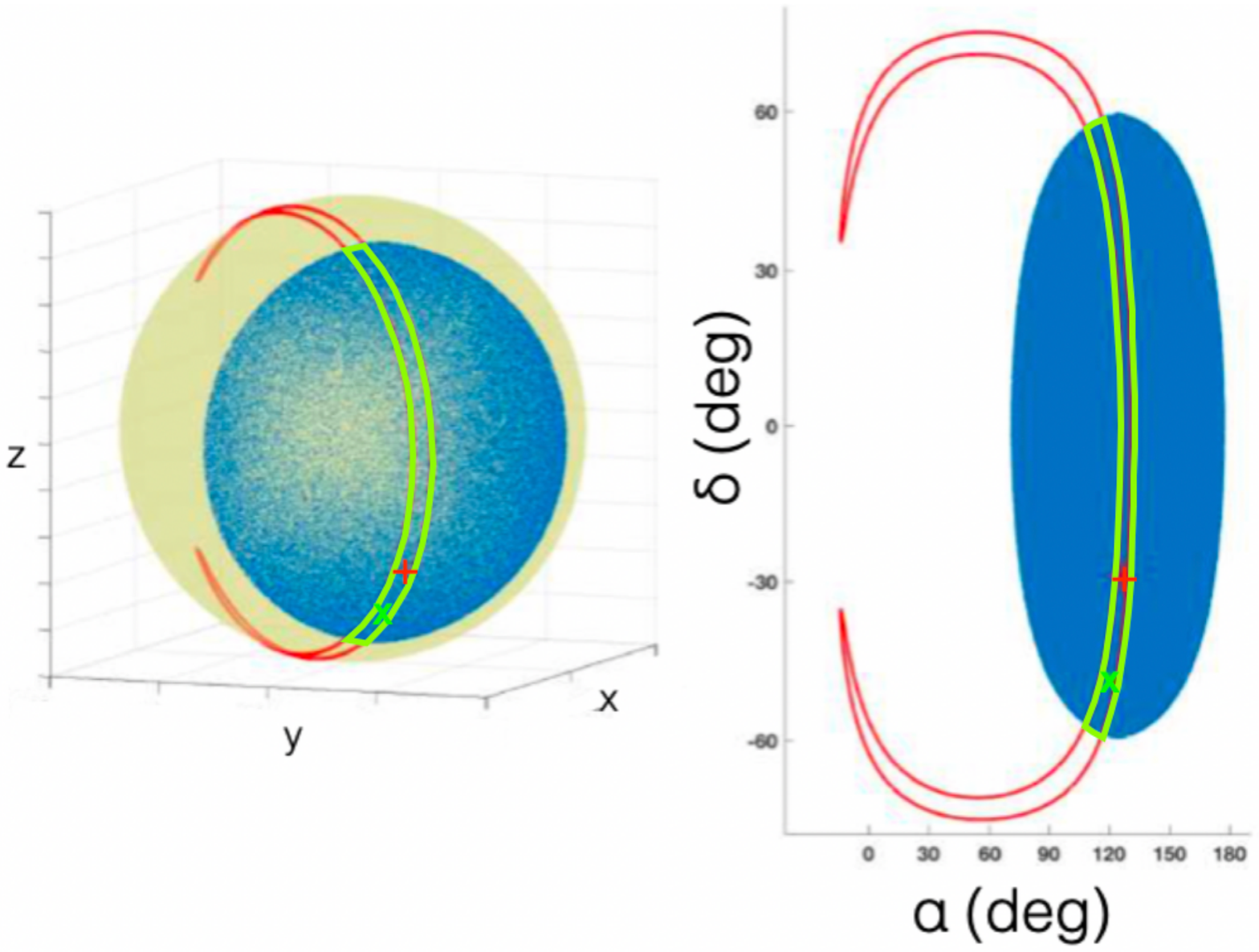} &
\includegraphics[scale=0.4,angle=90.0]{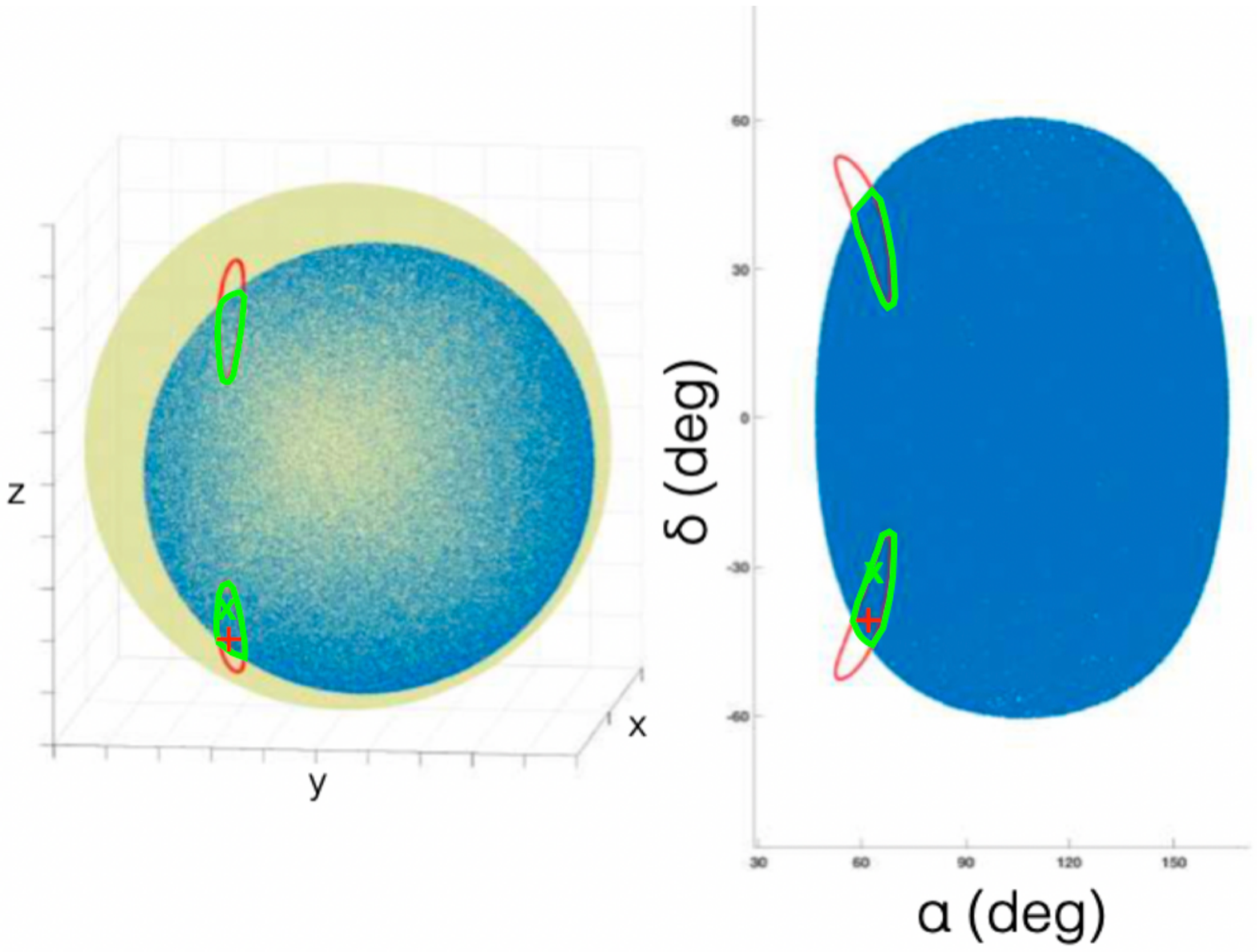}\\
\end{tabular}
\end{center}
\vspace{-3.5cm}
\caption[example] 
{ \label{fig:conf_reg} 
Example of GRB localization obtained by simulating a GRB with latitude < 30 deg (left panel) and latitude > 70 deg (right panel). The GRB position and best-fit position are represented with the green X-symbol and the red +-symbol, respectively.  The red contour line shows the 1$\sigma$ confidence level region associated with the best fit position. The blue-filled region represents the overlapping FoVs of the 3 satellites observing simultaneously the GRB. The light-green contour line shows the 1$\sigma$ confidence level region corrected to take into account the FoVs of the detectors.}
\end{figure}

In Figure~\ref{fig:conf_reg} we report an example of two possible outcomes obtained from the simulations described above. Left and right panels in the figure display the GRB real position in the plane of the sky (green X-symbol), the best-fit position of the GRB (red +symbol) and the corresponding $1\sigma$ confidence region shown both in spherical and bi-dimensional representations. Moreover, in light-green we report the positional $1\sigma$ confidence region corrected by overlapping FoVs of detectors (blue-filled region) observing simultaneously the event. Both these cases describe the localisation of GRB events detected simultaneously by 3 satellites. While for the event shown in the left panel of Figure~\ref{fig:conf_reg} the confidence region of the GRB coordinates is nearly unconstrained in the $\delta$ coordinate, the right panel describes a more favorable scenario in which the uncertainty on the GRB position is limited to a relatively small region. For the former case, we note that, although limited, two confidence regions are present, but only one includes the position of the GRB. This ambiguity reflects the planar symmetry of the signal delays with respect to the satellite orbital plane. Differences among these two cases are due to the superposition of several aspects such as the accuracy in the determination of the time delays, physical distances between the satellites and projected distances of the satellites with respect to the direction of the GRB event. Moreover, for the two cases discussed above (but also for the larger set of results obtained with the simulations) we note that the Right Ascension is always relatively well constrained, leaving the largest uncertainty in the Declination. This result is expected when considering triplets of satellites laying in the equatorial plane. Improvements in $\delta$ can be obtained by increasing the number of satellites simultaneously observing the GRB event and located in inclined plane with respect to the equatorial one.

\section{Results and Discussion}

We applied the method described in Sec.~\ref{sec:strat} for each of the 760 GRBs simulated within the 2-years time interval characterizing the lifetime of the HTP/HSP mission. To estimate confidence intervals of these quantities, we performed 1000 Monte Carlo simulations in which we followed the procedure extensively discussed earlier on. We note that by assuming the $\sigma_{cc-long}$ and $\sigma_{cc-short}$ values reported in Sec.~\ref{sec:grbcc} for all the simulated long and short GRBs, respectively, we are not correctly representing the GRB sample propertied. To mitigate this issue, the results reported below have been rescaled to take into account that $\sigma_{cc-long}$ and $\sigma_{cc-short}$ characterize only 55\% and 30\% of the long and the short GRB populations, respectively.

\subsection{Long GRBs}

\begin{figure}[h]
\centering
\includegraphics[scale=0.55,angle=0]{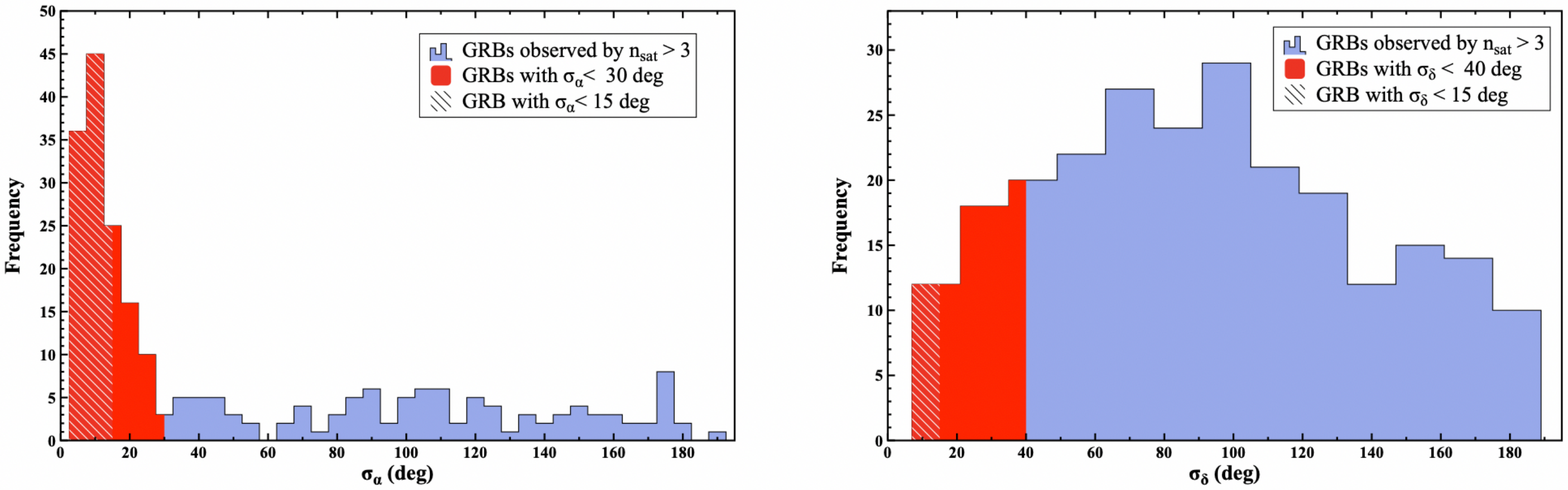}
\vspace{-3cm}
\caption{Distribution of the Right Ascension (left panel) and Declination (right panel) uncertainties from one of the simulations performed on the long GRBs. The red and white-hatched regions highlight the GRBs with $\alpha$ and $\delta$ uncertainties lower than $30^\circ$ ($40^\circ$ for $\delta$) and $15^\circ$, respectively.} 
\label{fig:res_long}
\end{figure}

Figure~\ref{fig:res_long} shows an example of the distribution of $\sigma_\alpha$ and $\sigma_\delta$ obtained from the observed GRBs within a 2-year long simulation. We emphasize that the number of observed GRBs reported in Figure~\ref{fig:res_long} needs to be rescaled by a factor of $\sim1.8$ to account for the assumed value of delay accuracy achievable $\sigma_{cc-long}$. In red and white (hatched region) we highlighted the detected GRBs for which the uncertainties on the two coordinates are lower or equal to $30^\circ$ ($40^\circ$ for $\delta$) and $15^\circ$, respectively. Figure~\ref{fig:res_long} clearly confirms the expected capabilities of the HTP/HSP set up, that allows us to be more sensitive to the Right Ascension with respect to the Declination coordinate.  

\vspace{-3cm}

Combining the results from the 1000 Monte-Carlo simulations, we obtained that among the 630 long GRBs generated in each simulation, $240\pm11$ are detected simultaneously by at least 3 detectors. We can further identify the following quantities:
\begin{itemize}
\item $66\pm5$ events with $\sigma_\alpha < 30^\circ$;
\item	$20\pm3$ events with $\sigma_\alpha < 30^\circ$ and $\sigma_\delta < 40^\circ$;
\item	$57\pm4$ events with $\sigma_\alpha < 15^\circ$;
\item	$16\pm2$ events with $\sigma_\alpha < 15^\circ$ and $\sigma_\delta < 40^\circ$;
\item	$7\pm2$ events with $\sigma_\alpha < 5^\circ$;
\item	$4\pm1$ events with $\sigma_\alpha < 5^\circ$ and $\sigma_\delta < 40^\circ$;
\item	$1\pm0.6$ events with $\sigma_\alpha < 5^\circ$ and $\sigma_\delta < 10^\circ$;
\end{itemize}

\subsection{Short GRBs}

\begin{figure}[h]
\centering
\includegraphics[scale=0.55,angle=0]{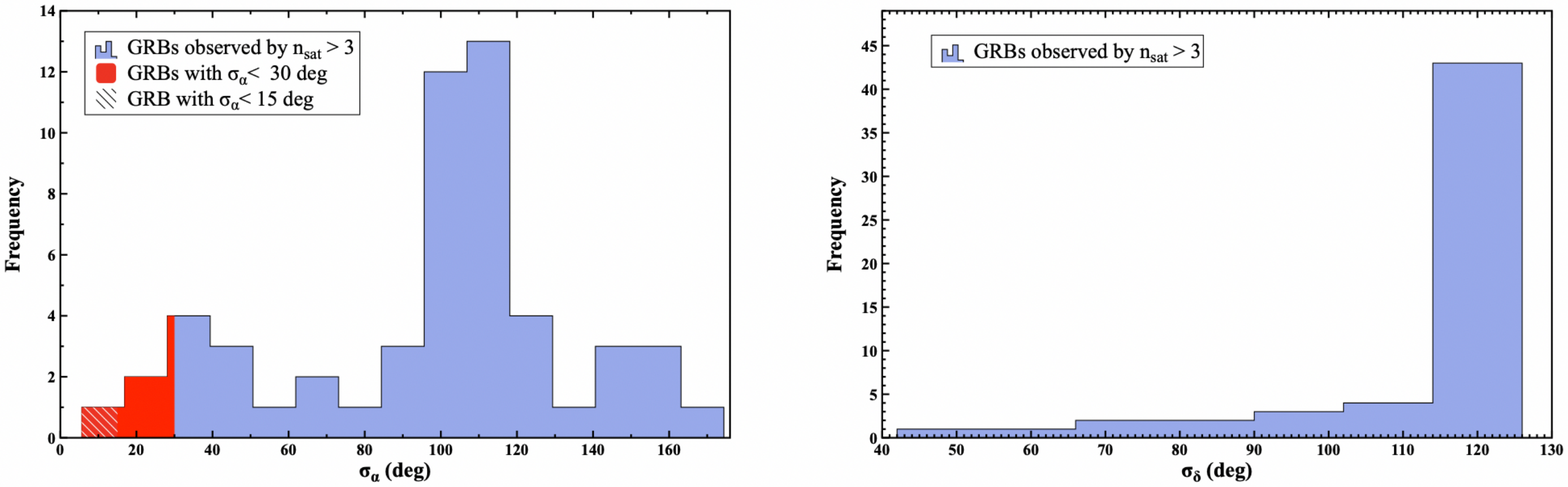}
\vspace{-3cm}
\caption{Distribution of the Right Ascension (left panel) and Declination (right panel) uncertainties from one of simulations of short GRBs. The red and white-hatched regions highlight the GRBs with $\alpha$ uncertainties lower than $30^\circ$ and $15^\circ$, respectively.} 
\label{fig:res_short}
\end{figure}

Figure~\ref{fig:res_short} shows the distribution of $\sigma_\alpha$ and $\sigma_\delta$ obtained from one simulation of the observed GRBs within a 2-year long mission (to be rescaled by a factor of $\sim3$ to consider the assumed value of $\sigma_{cc-short}$). Red and white (hatched) regions represent the detected GRBs for which the uncertainties on the $\alpha$ coordinate are lower or equal to $30^\circ$ and $15^\circ$, respectively. Combining the results from the 1000 Monte-Carlo simulations, we obtained that among the 130 short GRBs generated in each simulation, $48\pm5$ are detected simultaneously by at least 3 detectors. As clear from Figure~\ref{fig:res_short}, short GRBs are less likely to be precisely localized with the adopted pointing strategy. In fact, the combination of the relatively large cross-correlation uncertainty ($\sigma_{cc-short} = 5$ ms) and the projected baselines allows us to predict only $2\pm1$ short GRBs located with $\sigma_\alpha < 30^\circ$. In the following section we will investigate alternative approaches on the pointing strategy to improve the localisation capabilities for long and short GRBs.

\vspace{-3cm}

\subsection{Localisation capabilities vs. satellite baselines}
As discussed in Sec. \ref{sec:po_strat}, it is worth stressing that the pointing strategy LVLH aims at optimizing the overlapping FoV of the detectors composing a triplet of satellites, in order to guarantee the maximization of the number of GRBs potentially localizable. As clear from the results reported above, a large number of events observed simultaneously by 3 (or more) satellites, although desirable, does not guarantee an equivalently large number of GRBs well located in the sky. With that in mind, adjustments on the pointing strategy could be applied to improve the localisation capabilities of the HTP/HSP mini-constellation.
Combining Eq.~\ref{eq:delayexp} and Eq.~\ref{eq:chisq}, it is possible to deduce that the key elements to accurately locate an event in the sky are the accuracy in determining the time delay between the observations in different detectors as well as their projected distance with respect the observed event. In Sec.~\ref{sec:grbcc}, we extensively discussed the former aspect, while here we will focus on the latter by discussing different aspects of the performed simulations. We start by exploring possible correlations between the projected baseline, parameter that describes the relative positions of two satellites with respect a generic GRB direction in the sky, and the accuracy on the equatorial coordinates used to localise an event. Taking as a reference a triplet of satellites, we can define three different projected baselines, one of which will depend upon the others. As a reference parameter we consider the smallest of the projected baselines defined within the triplet, we then take the whole dataset of long and short simulated GRBs for a specific orbit and we correlate them against the minimum projected baselines. 

\begin{figure}[h]
\centering
\includegraphics[scale=0.55,angle=0]{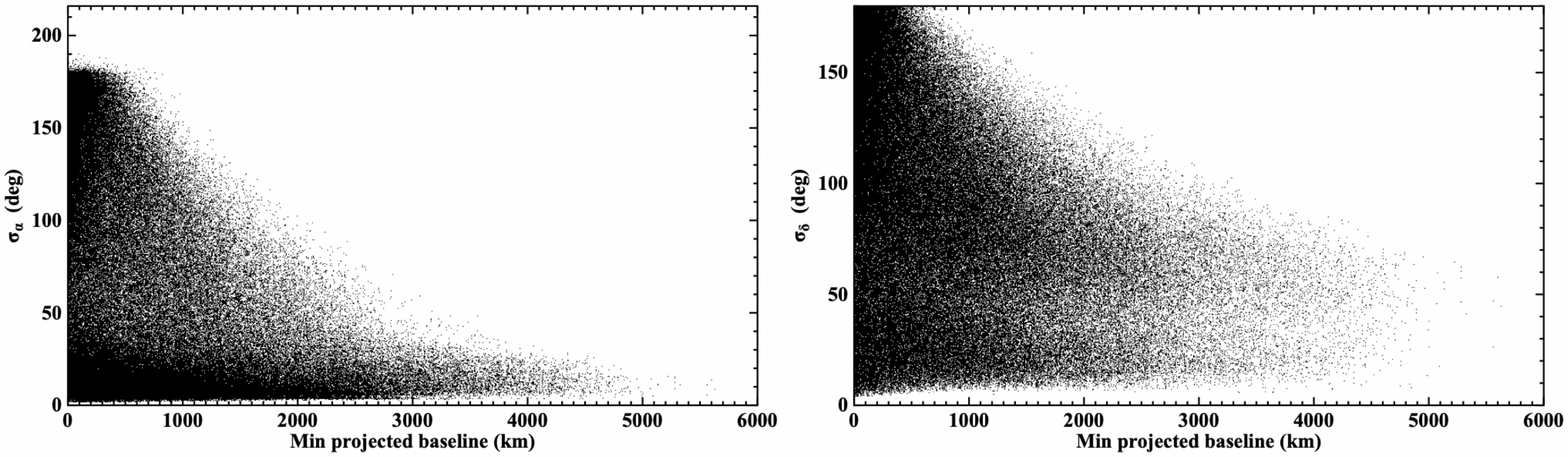}
\vspace{-3cm}
\caption{Uncertainties on the Right Ascension $\sigma_\alpha$ (left panel) and Declination $\sigma_\alpha$ (right panel) as a function of the minimum projected baseline within the triplet (or multiplet) of satellites that observed simultaneously the simulated long GRBs.} 
\label{fig:all_pos_long_sim}
\end{figure}

\begin{figure}[h]
\centering
\includegraphics[scale=0.55,angle=0]{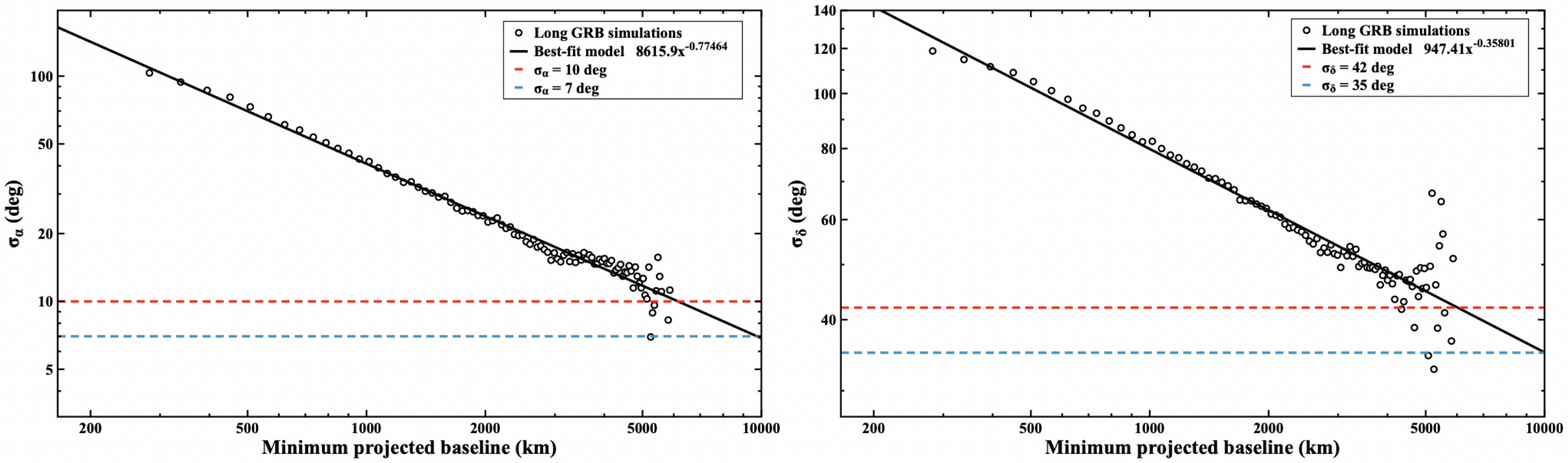}
\vspace{-3cm}
\caption{Average correlation between the uncertainties in the Right Ascension $\sigma_\alpha$ (left panel) and Declination $\sigma_\alpha$ (right panel) coordinates as a function of the minimum projected baseline obtained by binning the results from the simulated long GRB as detected by the HTP/HSP mini-constellation. The black-solid lines represent the best-fitting model to the data, whilst dashed red and cyan lines mark the achievable values of $\sigma_\alpha$ and $\sigma_\delta$ for minimum projected baselines of 6000 km and 10000 km, respectively.} 
\label{fig:all_pos_long_reb}
\end{figure}

Figure~\ref{fig:all_pos_long_sim} shows an example of the correlation between $\sigma_\alpha$ (left panel) and $\sigma_\delta$ (right panel) with respect to the minimum projected baseline for all the simulated long GRBs with localisation. The two plots clearly show a dispersion on the values $\sigma_\alpha$ and $\sigma_\delta$. Interestingly and in line with the predictions, this dispersion as well as the absolute value of the uncertainty significantly decrease at large values of the minimum projected baseline. To further investigate the correlations shown in Figure~\ref{fig:all_pos_long_sim}, we apply binning techniques to the datasets creating equally spaced intervals in the projected baseline parameter and averaging the values of $\sigma_\alpha$ and $\sigma_\delta$ within the selections. Figure~\ref{fig:all_pos_long_reb} (empty circles) shows the results of the binning method for the long GRB dataset of Figure~\ref{fig:all_pos_long_sim}. The figure describes the average localisation capabilities of the HTP/HSP configuration while observing randomly distributed long GRBs characterized by $\sigma_{cc-long} = 1$ ms. It is interesting to note that both the uncertainties on the equatorial coordinates decrease coherently with increasing values of the minimum projected baseline. To be able to quantify the variation of these parameters, we modelled the data with a generic power-law function. Based on the best-fit models (solid black lines in Figure~\ref{fig:all_pos_long_sim}), we can extrapolate the localisation capabilities of the mini-constellation at different values of the minimum projected baseline of a generic triplet defined within the LVLH pointing strategy. It is worth to notice that by operating the two triplets with an average minimum projected baseline of the order of 6000 km, the mini-constellation will be able to routinely locate all the long GRBs characterized by $\sigma_{cc-long} = 1$ ms (half of observable sample) with an accuracy $\sigma_\alpha \leq 10^\circ$ and $\sigma_\delta \leq 42^\circ$ (red-dashed lines). By increasing the minimum projected baseline up to 10000 km, the accuracies will become of the order of $\sigma_\alpha \leq 7^\circ$ and $\sigma_\alpha \leq 35^\circ$ (cyan-dashed lines). 

\begin{figure}[h]
\centering
\includegraphics[scale=0.55,angle=0]{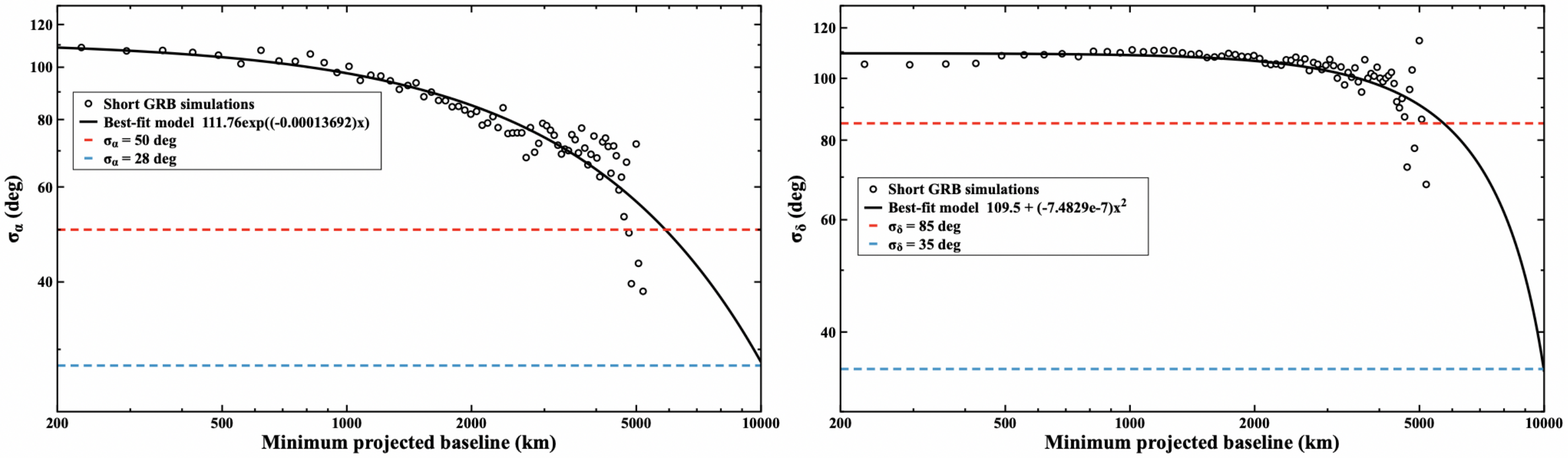}
\vspace{-3cm}
\caption{Average correlation between the uncertainties in the Right Ascension $\sigma_\alpha$ (left panel) and Declination $\sigma_\alpha$ (right panel) coordinates as a function of the minimum projected baseline obtained by binning the results from the simulated short GRB as detected by the HTP/HSP mini-constellation. The black-solid lines represent the best-fitting model to the data, whilst dashed red and cyan lines mark the achievable values of $\sigma_\alpha$ and $\sigma_\delta$ for minimum projected baselines of 6000 km and 10000 km, respectively.} 
\label{fig:all_pos_short_reb}
\end{figure}

We performed a similar analysis for the short GRBs. In analogy with the long ones, Figure~\ref{fig:all_pos_short_reb} (empty circles) represents the average localisation capabilities of the HTP/HSP configuration with respect to randomly distributed short GRBs characterized by $\sigma_{cc-short} = 5$ ms. Also in this case, both the uncertainties on the equatorial coordinates decrease coherently with increasing values of the minimum projected baseline, although the behavior at large value of the baseline starts to be more complex. To be able to quantify the variation of these parameters, we modelled the data with an \emph{ad hoc} function reported in the figure legend. Based on the best-fit models (solid black lines in figure), we can extrapolate the localisation capabilities of the mini-constellation at different values of the minimum projected baseline of a generic triplet. It is interesting to notice that by operating the two triplets with an average minimum projected baseline of the order of 6000 km it will be possible to routinely locate the short GRBs characterized by $\sigma_{cc-short} = 5$ ms (~30\% of the complete sample) with an accuracy $\sigma_\alpha \leq 50^\circ$ and $\sigma_\delta \leq 85^\circ$ (red-dashed lines). Increasing the minimum projected baseline up to 10000 km will allow to reduce the accuracies at values $\sigma_\alpha \leq 28^\circ$ and $\sigma_\delta \leq 35^\circ$ (cyan-dashed lines).

It is noteworthy that, from simple geometrical considerations, increasing the projected baseline of the satellites will imply the reduction of overlapping FoV of the detectors with a consequent decrease in number of GRBs detected. On the other hand, the fraction of events with good localisation will significantly increase. Therefore, a trade-off study will be carried out to define adjustments on the mission strategy able to maximize the scientific outcomes by defining a suitable number of accurately localized GRBs events.

\section{Summary}

In this paper, we characterized the localisation capabilities of the HTP/HSP mini-constellation based on the application of temporal triangulation methods on the observed GRBs. We extensively investigated the validity of cross-correlation techniques to accurately recover time delays between the GRB light-curves detected by multiple elements of constellation. By studying unbiased samples of 100 long and 100 short GRBs, we deeply studied correlations between GRB properties and cross-correlation capabilities to determine delays between light-curves. Finally, based on a specific mission scenario and an optimized pointing strategy for the detectors, we simulated the HTP/HSP performances during a 2-years lifetime mission by providing an estimate of the number of localized GRBs. Moreover, we estimated positional accuracies of the detected GRBs depending on their properties. Finally, we investigated variations of the observational set-up to improve the localisation capabilities of the mini-constellation. Based on the results obtained from these simulations, we can conclude that the HERMES Pathfinder will achieve its scientific goal to collected enough GRB detections and localizations to validate the HERMES overall concept, as well as proving crucial insights for the design of an extended version of the constellation.


\acknowledgments 
 
This work has been carried out in the framework of the HERMES-TP and HERMES-SP collaborations. We acknowledge support from the European Union Horizon 2020 Research and Innovation Framework Programme under grant agreement HERMES-Scientific Pathfinder n. 821896 and from ASI-INAF Accordo Attuativo HERMES Technologic Pathfinder n. 2018-10-H.1-2020.  

\bibliography{report}
\bibliographystyle{spiebib} 

\end{document}